\documentclass[%
reprint,
superscriptaddress,
 amsmath,amssymb,
 aps,
prx,
floatfix,
]{revtex4-1}
\usepackage{graphicx}
\usepackage{hyperref}
\usepackage[utf8]{inputenc} 
\usepackage[T1]{fontenc}
\usepackage[usenames,dvipsnames]{xcolor}
\usepackage{braket}
\usepackage{ bbold }

\DeclareMathOperator{\Tr}{Tr}

\definecolor{blue(pigment)}{rgb}{0.2, 0.2, 0.6}
\definecolor{darkerblue}{rgb}{0.0, 0.0, 0.4}
\definecolor{darkblue}{rgb}{0.0,0.0,0.5}
\definecolor{darkgreen}{rgb}{0.0,0.4,0.0}
\hypersetup{
   colorlinks,
   linkcolor=blue(pigment),
    citecolor=darkgreen,
   urlcolor=darkblue
}

\begin{document}

\title{Probing eigenstate thermalization with the emergence of fluctuation-dissipation relations in quantum simulators}%
\title{Probing eigenstate thermalization in quantum simulators via fluctuation-dissipation relations}%
\author{Alexander Schuckert}\email{alexander.schuckert@tum.de}
\affiliation{Department of Physics and Institute for Advanced Study, Technical University of Munich, 85748 Garching, Germany}
\affiliation{Munich Center for Quantum Science and Technology (MCQST), Schellingstr. 4, D-80799 M\"unchen}
\author{Michael Knap}
\email{michael.knap@ph.tum.de}
\affiliation{Department of Physics and Institute for Advanced Study, Technical University of Munich, 85748 Garching, Germany}
\affiliation{Munich Center for Quantum Science and Technology (MCQST), Schellingstr. 4, D-80799 M\"unchen}
\date{\today}

\begin{abstract}
The eigenstate thermalization hypothesis (ETH) offers a universal mechanism for the approach to equilibrium of closed quantum many-body systems. So far, however, experimental studies have focused on the relaxation dynamics of observables as described by the diagonal part of ETH, whose verification requires substantial numerical input. This leaves many of the general assumptions of ETH untested. Here, we propose a theory-independent route to probe the full ETH in quantum simulators by observing the emergence of fluctuation-dissipation relations, which directly probe the off-diagonal part of ETH. We discuss and propose protocols to independently measure fluctuations and dissipations as well as higher-order time ordered correlation functions. We first show how the emergence of fluctuation dissipation relations from a nonequilibrium initial state can be observed for the 2D Bose-Hubbard model in superconducting qubits or quantum gas microscopes. Then we focus on the long-range transverse field Ising model (LTFI), which can be realized with trapped ions. The LTFI exhibits rich thermalization phenomena: For strong transverse fields, we observe prethermalization to an effective magnetization-conserving Hamiltonian in the fluctuation dissipation relations. For weak transverse fields, confined excitations lead to non-thermal features resulting in a violation of the fluctuation-dissipation relations up to long times. Moreover, in an integrable region of the LTFI, thermalization to a generalized Gibbs ensemble occurs and the fluctuation-dissipation relations enable an experimental diagonalization of the Hamiltonian. Our work presents a theory-independent way to characterize thermalization in quantum simulators and paves the way to quantum simulate condensed matter pump-probe experiments.
\end{abstract}

\maketitle
\section{Introduction}
The long coherence time scales accessible in quantum simulators made it possible to experimentally observe thermalization in isolated quantum systems~\cite{trotzky_probing_2012,PhysRevLett.113.147205,kaufman_quantum_2016,PhysRevX.8.021030}, the absence thereof in the presence of disorder~\cite{schreiber_observation_2015, smith_many-body_2016,choi_exploring_2016, PhysRevX.7.041047,brydges_probing_2019} and integrability in reduced dimensions~\cite{kinoshita_quantum_2006,gring_relaxation_2012}. Typically, these observations were based on probing \emph{equal-time} correlation functions~\cite{schweigler_experimental_2017, zache_extracting_2019,prufer_experimental_2019}, concluding the observation of equilibration by comparison to the expected microcanonical expectation values at the same energy density as the initial state. This approach in particular requires viable theory input to compare with. However, in order to show full thermalization also the \emph{fluctuations} around the equilibrium expectation value as well as the \emph{response} of the system to small perturbations need to match the expectation in thermal equilibrium. This can be understood from the eigenstate thermalization hypothesis (ETH)~\cite{deutsch_quantum_1991,srednicki_chaos_1994,srednicki_approach_1999,rigol_thermalization_2008,reimann_foundation_2008}, via its Ansatz for the matrix elements of observables $\hat A$ with respect to many-body eigenstates $\ket{n}$ with energy $E_n$:
\begin{equation}
\braket{n|\hat A|m} = A(\bar E)\delta_{nm} + e^{-S(\bar{E})/2}f_A(\bar E,\omega)R_{nm},
\end{equation}
where $\bar{E}=(E_n+E_m)/2$, $\omega=E_m-E_n$, $A(\bar E)$ is the value of $\langle\hat A\rangle$ in the microcanonical ensemble at energy $\bar E$, $S(\bar E)$ is the thermodynamic entropy (i.e. the number of states in a small interval around energy $\bar E$) and $R_{nm}$ are Gaussian random numbers. Measuring equal-time correlation functions in experiment only probes the first (``diagonal'') term as in the long time limit $\braket{\hat A(t)}\equiv\braket{\psi_0|\hat A(t)|\psi_0} \rightarrow \bar A\equiv \sum_n |\braket{\psi_0|n}|^2 \braket{n|\hat A|n}$. While temporal fluctuations of equal-time correlation functions around the steady-state value can in principle be used to probe the off-diagonal part of ETH as $\braket{\hat A(t)}^2-\bar A^2\rightarrow \sum_{m\neq n}|\braket{\psi_0|n}|^2|\braket{m|\psi_0}|^2|\braket{m|\hat A|n}|^2$, they are exponentially small in system size since the thermodynamic entropy is extensive. Hence, it becomes impractical to observe them in large systems~\cite{PhysRevE.99.052139,kaplan_many-body_2020}. Equal-time correlation functions therefore only probe the diagonal part of ETH while requiring substantial theory input to conclude thermalization in experiment as they require a comparison with an equilibrium expectation value.

\begin{figure}
	\includegraphics{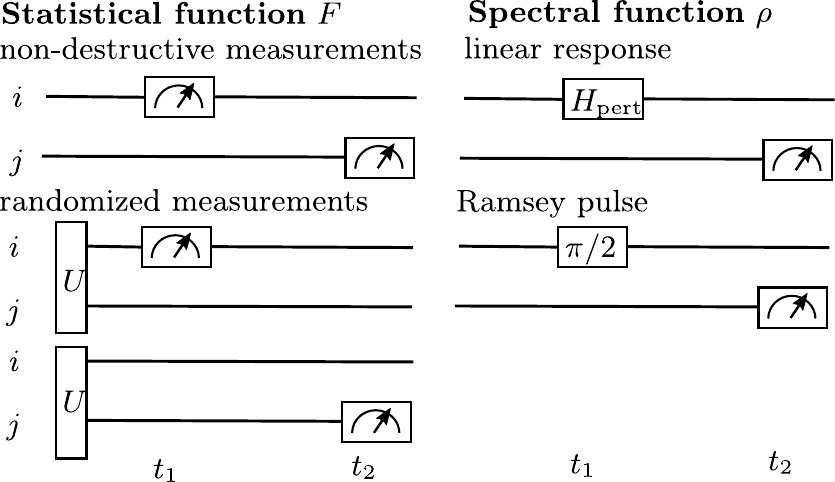}
	\caption{\textbf{Measuring two-time correlation functions out of equilibrium.} The statistical function $F$ can be measured by employing a non-destructive measurement on site $i$ at time $t_1$ before measuring site $j$ at time $t_2$. The measurement at $t_1$ can be deferred to $t_2$ by shelving. Alternatively, measurements of two independent experimental realizations can be combined to yield $F$ by averaging over global random unitaries $U$ acted on the initial state. The spectral function $\rho$ can be measured by non-equilibrium linear-response (e.g. Bragg or tweezer spectroscopy), employing light pulses on lattice site $i$ at time $t_1$ before measuring at time $t_2$. Alternatively, a Ramsey-type sequence works similarly. The protocols for $F$ and $\rho$ can be realized in quantum simulators of spin models such as trapped ion experiments as well as simulators of Bose- and Fermi-Hubbard models such as quantum gas microscopes and superconducting qubits. The non-destructive measurement and Ramsey protocols can be combined to measure higher-order time ordered correlation functions.\label{fig:measurement}}
\end{figure}

Here, we propose to measure \emph{two-time correlation functions} of the form $\braket{\hat A(t_1) \hat B(t_2)}$ to probe thermalization in quantum simulators. They are entirely determined by the off-diagonal part of ETH while staying of $\mathcal{O}(1)$ in the thermodynamic limit, hence offering a route to experimentally probe the entirety of eigenstate thermalization.
Moreover, two-time correlation functions offer  a completely \emph{theory-independent} route to do so by testing the fluctuation-dissipation relations (FDRs)~\cite{PhysRev.83.34,berges_thermalization_2001, babadi_far--equilibrium_2015, PhysRevLett.122.150401,khatami_fluctuation-dissipation_2013,randi_probing_2017}.  FDRs relate the anticommutator (statistical) two-time function  
\begin{equation}
F(t_1,t_2)=\frac{1}{2}\braket{\lbrace\hat A(t_1),\hat B(t_2)\rbrace}-\braket{\hat A(t_1)}\braket{\hat B(t_2)},
\label{eq:F}
\end{equation}
which quantifies fluctuations of the system, with the commutator (spectral function)
\begin{equation}
\rho(t_1,t_2)=\braket{[\hat A(t_1),\hat B(t_2)]},
\label{eq:Rho}
\end{equation}
quantifying dissipation of energy~\footnote{Heating rates in the linear response regime of periodically driven systems are determined by $\rho$~\cite{PhysRevLett.124.106401} and hence can also be used to probe the off-diagonal part of ETH~\cite{PhysRevLett.123.240603}. However, heating rates are challenging to measure in experiment~\cite{PhysRevX.10.021044}.}. Once local thermal equilibrium is approached at late times, the fluctuations and dissipations are independent of the central time $T=(t_1+t_2)/2$ due to time-translational invariance. Fourier transforming the relative time $\tau=(t_1-t_2)$ to frequencies $\omega$, we obtain the FDR
\begin{equation}
F(\omega)=n_\beta(\omega) \rho(\omega).
\label{eq:FDR}
\end{equation}
The Bose-Einstein distribution (plus the ``quantum half'') $n_\beta(\omega)=1/2+1/(\exp(\beta\omega)- 1)$ at inverse temperature $\beta$ links fluctuations and dissipation; see App.~\ref{app:KMS_FDR} for a short derivation of the FDRs and App.~\ref{app:ETH_FDR} for the connection between ETH and FDRs. As the FDR is completely independent of microscopic details and the initial state, measuring $F$ and $\rho$ independently from each other out-of-equilibrium and testing the FDRs provides a universal and theory-independent way of probing thermalization in quantum simulators. Moreover, from the FDR one can extract the temperature of the many-body system, which is usually challenging to determine experimentally~\cite{PhysRevLett.106.225301,hartke_measuring_2020}. 

While the ETH implies the fulfillment of FDRs, physical initial states are always superpositions of many eigenstates such that the FDR of single eigenstates are challenging to probe in experiments~\footnote{A recent protocol has shown how to prepare eigenstates in finite size systems~\cite{yang_quantum_2020}.}. However, the energy density variance of initial states prepared as the ground state of some Hamiltonian can be shown to vanish in the thermodynamic limit~\cite{rigol_thermalization_2008}. Hence, such initial states can be seen as a superposition of eigenstates in a small energy shell. As the ETH functions $A(\bar E)$ and $f_A(\bar E,\omega)$ are assumed to be smooth functions of $\bar E$, the vanishingly small energy variance of physical initial state implies that probing an initial state with energy $\bar E$ and probing an arbitrary eigenstate with the same energy yields the same result at long times.

In this work, we propose protocols for measuring fluctuations and dissipations independently from each other out-of-equilibrium in quantum simulators of spin systems as well as fermionic and bosonic quantum gas microscopes employing protocols based on Ramsey pulses~\cite{knap_probing_2013}, non-destructive projective measurements~\cite{knap_probing_2013, uhrich_noninvasive_2017}, randomized measurements~\cite{PhysRevLett.120.050406} and linear response, including non-equilibrium Bragg~\cite{rey_bragg_2005} and ``tweezer'' spectroscopy (Sec.~\ref{sct:measurement}). We then discuss applications of the protocols in Sec.~\ref{sct:appli}. As a first example, we show that the FDRs can be probed in current quantum gas microscopes as well as superconducting qubit experiments implementing the Bose-Hubbard model. Going beyond the case of fast thermalization, we show that in trapped ion experiments several examples of prethermalization~\cite{berges_prethermalization_2004} can be probed in the long-range transverse field Ising model (LTFI). At large transverse fields, a single approximately conserved quantity leads to thermalization to a prethermal Hamiltonian, which can be directly observed by testing the FDRs. In an integrable sector of the LTFI, extensively many conserved quantities lead to thermalization to a generalized Gibbs ensemble, which can again be observed by a generalized FDR~\cite{PhysRevE.95.052116}. In turn, measuring two-time correlations enable an experimental diagonalization of the quadratic Hamiltonian. Finally, at small transverse fields, confined excitations can be directly observed in the spectral function and lead to genuine non-thermal features including a violation of the FDR observable up to long times.

\section{Measuring n-time correlation functions in quantum simulators\label{sct:measurement}}
Solving the quantum many-body problem is equivalent to obtaining all time ordered correlation functions~\cite{schweigler_experimental_2017} $\langle T \hat A(t_1) \hat B(t_2) \hat C(t_3)\cdots \rangle$. Here, we propose protocols to measure such correlation functions in quantum simulators of lattice models by using their decomposition into nested (anti-) commutators~\cite{chou_equilibrium_1985}. In particular, we will focus on the two-time correlation function which can be decomposed into the anti-/commutator (i.e. the statistical/spectral function) according to $\langle T \hat A(t_1) \hat B(t_2)\rangle=F+\frac{1}{2}\mathrm{sgn}(t_1-t_2)\rho$. In the following, we present several protocols to measure $F$ and $\rho$ independently from each other in quantum simulators of spin and Bose-/Fermi-Hubbard models and indicate how to generalize them to higher order time ordered correlation functions. Our protocols are summarized in Fig.~\ref{fig:measurement}.

\subsection{Simulators of spin models}
\subsubsection*{Ramsey protocol for spectral function $\rho$} many-body Ramsey interferometry has been shown to be probe spectral functions in quantum simulators of spin models~\cite{knap_probing_2013, uhrich_noninvasive_2017} using local spin rotations of the form
\begin{equation}
R_i^\alpha(\theta)=\cos(\theta/2)\hat{\mathbb{1}}_i-i\sin(\theta/2)\hat{\sigma}^\alpha_i,
\end{equation}
where $\hat \sigma^\alpha$ are the Pauli matrices~\footnote{While Rabi pulses only directly implement pulses in the x-y plane of the Bloch sphere, a pulse around the z axis can be implemented by $\hat{R}^z_i(\theta)=\hat{R}^x_i(\pi/2)\hat{R}^y_i(\theta)\hat{R}^x_i(-\pi/2)$~\cite{haffner_quantum_2008}.}. The protocol proceeds as follows: Starting from some initial state $\ket{\Psi_0}$, evolve for time $t_1$, apply a local rotation $R_i^\alpha(\theta)$ at site $i$, subsequently evolve for a time ($t_2-t_1$) and finally measure $\hat{\sigma}_j^\beta$~\footnote{Measurements of $\hat{\sigma}^{x/y}$ can be implemented by applying local pulses before measuring $\hat \sigma^z$, for example $\hat{\sigma}^y=-\hat{R}^x(-\frac{\pi}{2})\hat{\sigma}^z\hat{R}^x(\frac{\pi}{2})$.}. The result can be written as
\begin{align}
\braket{\hat{\sigma}_j^\beta(t_2)}_\theta&=\cos^2(\theta/2)\braket{\hat{\sigma}_j^\beta(t_2)}+\frac{i}{2}\sin{\theta}\braket{[\hat{\sigma}_i^\alpha(t_1),\hat{\sigma}_j^\beta(t_2)]}\notag\\&+\sin^2(\theta/2)\braket{\hat{\sigma}_i^\alpha(t_1)\hat{\sigma}_j^\beta(t_2)\hat{\sigma}_i^\alpha(t_1)},
\label{eq:comm_exp}
\end{align}
where all expectation values are written in the Heisenberg picture. The spectral function can then be obtained by combining two runs with opposite angle $\theta=\pm \pi/2$ by
\begin{equation}
\braket{[\hat{\sigma}_i^\alpha(t_1),\hat{\sigma}_j^\beta(t_2)]}=-i\braket{\hat{\sigma}_j^\beta(t_2)}_{\pi/2}+i\braket{\hat{\sigma}_j^\beta(t_2)}_{-\pi/2}.
\end{equation}

\subsubsection*{Projective measurement protocol for $F$} The statistical function $F$ has been shown to be probed by replacing the pulses in the Ramsey protocol for $\rho$ with non-destructive projective measurements~\cite{knap_probing_2013,uhrich_noninvasive_2017}, which have for example been demonstrated in superconducting qubits~\cite{lupascu_quantum_2007}, Rydberg tweezer arrays~\cite{PhysRevLett.122.173201} and trapped ions~\cite{PhysRevLett.56.2797,PhysRevLett.57.1696}. In this protocol, a measurement of $\hat{\sigma}_i^\alpha$ at time $t_1$ (without disturbing the rest of the system) and a subsequent measurement of $\hat{\sigma}_j^\beta$ at time $t_2-t_1$ is combined to yield 
\begin{equation}
\frac{1}{2}\braket{\big\lbrace\hat{\sigma}_i^\alpha(t_1),\hat{\sigma}_j^\beta(t_2)\big\rbrace} = P^{+\alpha+\beta}_{ij}+P^{-\alpha-\beta}_{ij}-P^{+\alpha-\beta}_{ij}-P^{-\alpha+\beta}_{ij},
\end{equation}
where $P^{+\alpha+\beta}_{ij}$ is the joint probability of measuring $+1$ for $\hat{\sigma}_i^\alpha(t_1)$ and $+1$ for $\hat{\sigma}_j^\beta(t_2)$.

The non-desctructive projective measurement can be replaced by spin shelving as noted in Ref.~\cite{knap_probing_2013}. In this variant, the measurement at time $t_1$ is replaced by a $\pi$ pulse between one of two spin levels at site $i$ and a third level, which does not participate in the many-body dynamics. At time $t_2$ this third level gets measured as well, effectively projecting the state onto one of the two measurement outcomes at time $t_1$. This variant of the protocol has a speed advantage as single-site pulses are usually much faster than measurements and the many-body dynamics. 

\subsubsection*{Randomized measurement protocol for $F$} We propose statistical correlations between randomized measurements~\cite{PhysRevLett.108.110503,ohliger_efficient_2013,elben_renyi_2018,PhysRevLett.122.120505,brydges_probing_2019,PhysRevX.9.021061} as an alternative to measure the statistical correlation function $F$ in small systems. It relies on acting with global random unitaries $\hat u$ on the initial state $\ket{\Psi_0}$. After time evolving for a time $t_1$,  $\hat A$ is measured. Preparing the same initial state (with the same unitary $\hat u$) to measure $\hat B$ after evolving for time $t_2$ as well as measuring the overlap $\braket{\rho_0}_u \equiv |\braket{\Psi_0|\hat u|\Psi_0}|^2$ of the initial state with $\hat u\ket{\Psi_0}$ in a separate measurement, one can then extract $F$ by averaging over random unitaries as
\begin{equation}
\braket{\lbrace \hat A,\hat B\rbrace}=\mathcal{N}_H^3\overline{\braket{\hat A(t_1)}_u \braket{\hat B(t_2)}_u \braket{\rho_0}_u}-\mathcal{N}_H C(t_1,t_2), \label{eq:F_noise}
\end{equation}where $\mathcal{N}_H\gg 1$ is the Hilbert space dimension, the overline denotes averaging over random unitaries and we assumed $\hat A$, $\hat B$ to be traceless. The second term is the infinite temperature correlation function
\begin{align}
C(t_1,t_2) &\equiv \frac{1}{\mathcal{N}_H}\Tr(\hat A(t_1)\hat B(t_2))\notag\\
&=\mathcal{N}_H\overline{\braket{\hat A(t_1)}_u \braket{\hat B(t_2)}_u},\label{eq:C_noise}
\end{align}
which is interesting in its own right as it quantifies thermalization and transport in the middle of the spectrum in systems with a bounded local Hilbert space. Note that both $F$ and $C$ can be obtained from the same experimental data. Moreover, if $\hat{A}=\hat{B}$ only a \emph{single} time trace needs to be measured for every unitary $u$ (along with $\braket{\rho_0}_u$). We present the proofs of Eqs. \eqref{eq:F_noise} and \eqref{eq:C_noise}, in App.~\ref{app:rand}, along with a generalization to operators which are not traceless and a simplification of the protocol in case of thermal equilibrium $\hat \rho_0\propto e^{-\beta \hat H}$. In the proofs, we assume $u$ to be a unitary 3-design, i.e. moments up to the third order have to match the circular unitary ensemble ($C$ can be measured with 2-designs). 

Global random unitaries can be implemented by adding local quenched disorder to a many-body Hamiltonian~\cite{PhysRevX.7.021006,PhysRevLett.120.050406}.

\subsubsection*{Higher order time ordered correlation functions}
Here, we generalize the previously known protocols for two-time functions~\cite{knap_probing_2013,uhrich_noninvasive_2017} to multi-time correlation functions. A specific three-point correlation function can be directly read off of Eq.~\eqref{eq:comm_exp}:
\begin{equation}
\braket{\hat{\sigma}_j^\beta(t_2)}_{\pi/2}+\braket{\hat{\sigma}_j^\beta(t_2)}_{-\pi/2} = \braket{\hat{\sigma}_i^\alpha(t_1)\hat{\sigma}_j^\beta(t_2)\hat{\sigma}_i^\alpha(t_1)},
\end{equation}
with $t_2>t_1$ as demanded by causality. In order to reconstruct the complete three point time ordered correlation function, we need to additionally measure all possible (anti-)commutator nestings~\cite{chou_equilibrium_1985}. These can be obtained by combining the projective measurement and Ramsey protocols as we show in App.~\ref{eq:threepoint}. For example, a measurement of $\hat{\sigma}^\alpha_i$ at time $t_1$ followed by a pulse $R_j^\beta(\theta)$ at time $t_2$ and a measurement of $\hat{\sigma}^\gamma_k$ at time $t_3$ can be combined to obtain 
\begin{align}
&\bra{\Psi(t_1)}\hat{P}_i^{+\alpha}\ket{\Psi(t_1)}\braket{\hat{\sigma}_j^\beta(t_2)}_{+\alpha,\theta=\pi/4}\notag \\&+\bra{\Psi(t_1)}\hat{P}_i^{-\alpha}\ket{\Psi(t_1)}\braket{\hat{\sigma}_j^\beta(t_3)}_{-\alpha,\theta=-\pi/4}\notag\\&=\frac{1}{4}\braket{\lbrace\hat{\sigma}_i^\alpha(t_1),[\hat{\sigma}_k^\gamma(t_2),\hat{\sigma}_j^\beta(t_3)]\rbrace},
\end{align} 
where $\hat{P}_i^{\pm \alpha} = \frac{1}{2}(\hat{\mathbb{1}}\pm\hat{\sigma}_i^{\pm \alpha})$ is the projection operator corresponding to eigenvalue +1/-1 of $\hat \sigma^\alpha$~\footnote{While projections on $\hat \sigma^z$ are directly implemented by measurements of the level population, projection operators in the other directions of the Bloch sphere may be implemented by precluding the measurement with appropriate pulses, $\hat{P}^{\pm y}=\hat{R}^x(\pi/2)\hat{P}^{\pm z}\hat{R}^x(-\pi/2)$ and $\hat{P}^{\pm x}=\hat{R}^y(\pi/2)\hat{P}^{\pm z}\hat{R}^y(-\pi/2)$.}. As we see above, a projective measurement/pulse results in the appearance of an anticommutator/commutator. We hence argue that this procedure generalizes to all n-point time ordered correlation functions by decomposing them into nested anti-/commutators.

\subsection{Simulators of Bose- and Fermi-Hubbard models}
By generalizing the previously discussed protocols for spin systems we show how to measure n-time correlation functions of the local density operator $\hat n_i$ in quantum simulators of bosonic or fermionic lattice models.
\subsubsection*{Ramsey protocol for spectral function $\rho$} A pulse operator $\hat R_i(\theta)$ analogous to the spin model protocol can be introduced by noting that the local density operator can be written as $\hat n_i=(\sigma^z_i-\mathbb{1})/2$ if the occupations are restricted to zero and one as the case for fermions and ``hard-core'' bosons, i.e. bosons in the presence of large on-site interactions. An off-resonant light field induces an AC Stark shift described by the Hamiltonian $\hat{H}_{\mathrm{L}}=-h_i \hat n_i$, which in a quantum gas microscope can be implemented by a ``tweezer'' laser shone on a single lattice site $i$, for example through a spatial light modulator~\cite{choi_exploring_2016}. In a superconducting circuit, this Hamiltonian can be implemented by a change in the frequency detuning of the superconducting oscillator representing lattice site $i$~\cite{roushan_spectroscopic_2017,chiaro_growth_2019}. In any case, applying the field for a duration $t$ implements the operator
\begin{equation}
\hat R_j(\theta)=\left(\cos(\theta/2)\mathbb{1}+i\sin(\theta/2)\hat \sigma_j^z\right)\exp(i\theta/2),
\end{equation} 
with $\theta=h_j t$ and we assumed $\hat H_L$ to be dominating the dynamics during the pulse. Proceeding as in the spin system protocol, i.e. evolving until time $t_1$, applying $R_i(\theta)$, evolving for a time $(t_2-t_1)$ and measuring $\hat n_j$, we get 
 \begin{align}
&\braket{\hat n_j(t_2)}_\theta=\braket{\hat n_j(t_2)}-i\sin(\theta)\braket{\left[\hat n_i(t_1),\hat n_j(t_2)\right]}\notag\\&+2\sin^2(\theta/2) \left[2\braket{\hat n_i(t_1)\hat n_j(t_2)\hat n_i(t_1)}-\braket{\left\lbrace\hat n_i(t_1),\hat n_j(t_2)\right\rbrace}\right],
\label{eq:nn_Rams}
\end{align} 
from which the spectral function can be extracted by choosing $\theta=\pm \pi/2$,
\begin{align}
\braket{\left[\hat n_k(t_1),\hat n_j(t_2)\right]}=\frac{i}{2}\left(\braket{\hat n_j(t_2)}_{\pi/2}-\braket{\hat n_j(t_2)}_{-\pi/2}\right).
\end{align}

\subsubsection*{Non-equilibrium linear response protocols for spectral function $\rho$} In non-equilibrium linear response, the spectral function may be obtained without restrictions on the occupation numbers. Here, we apply a small perturbation $\hat V$ during the dynamics and compare the measurement of an observable $\hat A$ at time $t_1$ to an evolution without perturbation. In general, the outcome of such an experiment is
\begin{equation}
\braket{\hat A(t_1)}_{V\neq 0} - \braket{\hat A(t_1)}_{V=0} = -i \int_{t_0}^{t_1} dt \braket{\left[\hat A(t_1), \hat V(t)\right]}.
\label{eq:response1}
\end{equation}
We now specify this expression to a local (real-space) and non-local (momentum-space) density perturbation.

\paragraph*{Local density perturbation.--} Applying a short pulse (compared to the many-body dynamics) with an off-resonant light field on lattice site $j$ such that $\hat V(t)=h_j \hat n_j \delta(t-t_2)$, where $h_j$ is the pulse area, we can measure the real space density-density spectral function via
\begin{equation}
\braket{\left[\hat n_k(t_1), \hat n_j(t_2)\right]}=\frac{i}{h_j}\left(\braket{\hat n_k(t_1)}_{h\neq 0} - \braket{\hat n_k(t_1)}_{h=0} \right),
\label{eq:response2}
\end{equation}
where $t_1>t_2$ due to causality and contrary to the Ramsey protocol, $h_j$ needs to be much smaller than the parameters of the many-body Hamiltonian. 

In the above protocol, separate experimental runs for different sites $j$ need to be conducted. By contrast, we can evaluate all $j$ simultaneously using a disordered global perturbation $\hat V(t)= \delta(t-t_2)\sum_k h_k \hat n_k $~\cite{PhysRevLett.122.150401}, with $\overline{h_i}=0$ and $\overline{h_i h_k}=\sigma_h^2\delta_{ik}$, where the overline denotes averaging over realizations of the random potentials with variance $\sigma_h^2$. The local spectral function can then be evaluated by post-processing as
\begin{align}
\braket{\left[\hat n_k(t_1), \hat n_j(t_2)\right]} =\frac{i}{\sigma_h^2}\overline{h_j\left(\braket{\hat n_k(t_1)}_{h\neq 0} - \braket{\hat n_k(t_1)}_{h=0}\right)},
\end{align}
where $\sigma_h^2$ needs to be small in order to be in the linear response regime.

\paragraph*{Stimulated Bragg spectroscopy.--} In Bragg spectroscopy~\cite{stamper-kurn_excitation_1999,stenger_bragg_1999,PhysRevA.67.053609,rey_bragg_2005}, two lasers are shone onto the lattice, with the atoms absorbing a photon from one of the two and emitting into the other. The momentum transfer $\hbar\mathbf{q}$ and the energy $\hbar \omega$ are defined by the angle between the two lasers and their frequency difference, respectively. The coupling to the atoms is given by
\begin{equation}
\hat V_I(t) = \frac{V_0}{2}\left( \hat n_\mathbf{-q}(t) e^{-i\omega t}+\hat n_\mathbf{q}(t) e^{i\omega t}\right) s(t), 
\end{equation}
where $\hat n_\mathbf{q}=\sum_j e^{i\mathbf{q}\mathbf{r}_j} \hat n_j$ is the Fourier transform of the local occupation numbers (i.e. the particle-hole excitation annihilation operator), $V_0$ is proportional to the laser intensity and $s(t)$ is the pulse envelope function. We consider measuring $\hat n_\mathbf{q}$ by using a quantum gas microscope to measure the local occupation numbers $\hat  n_j$ and Fourier transforming afterwards. In the following, we specify this protocol to two pulse shapes $s(t)$. 

Assuming a delta-like pulse, $s(t_1-t_p)\sim \delta(t_1-t_p)$, we get
\begin{equation}
\braket{\hat n_\mathbf{q}(t)} - \braket{\hat n_\mathbf{q}(t)}_{V=0} = -\frac{iV_0}{2} \braket{\left[\hat n_\mathbf{q}(t), \hat n_\mathbf{-q}(t_p)+\hat n_\mathbf{q}(t_p)\right]},
\end{equation}
i.e. the analogous expression to Eq.~\eqref{eq:response2} in momentum space. The Bragg pulses duration can be much slower than typical tunneling times in optical lattices, such that the $\delta$-form of the pulse is valid~\cite{stenger_bragg_1999}.

For a constant pulse, $s(t_1)=1$, a  Laplace transform with respect to $t$ evaluated at the same frequency $\omega$ results in
\begin{align}
&\braket{\hat n_\mathbf{q}(\omega)} - \braket{\hat n_\mathbf{q}(\omega)}_{V=0} = \notag\\&-\frac{iV_0}{2} \int_0^\infty dt \int_0^t dt_1 \big\langle\big[\hat n_\mathbf{q}(t), \hat n_\mathbf{-q}(t_1)e^{-i\omega (t_1-t)}\notag\\&\qquad\qquad\qquad\qquad\qquad+\hat n_\mathbf{q}(t_1)e^{i\omega (t_1+t)}\big]\big\rangle,
\end{align}
which is related to the spectral function Fourier transformed with respect to the relative time.
\subsubsection*{Projective measurement protocol for $F$}
 The projective measurement protocol for spin systems crucially relies on the fact that spin operators have exactly two eigenvalues. In simulators of Fermi-Hubbard models, the spin system protocol can therefore be straightforwardly generalized to the measurement of the local density $\hat n_{i\sigma}$ of hyperfine/spin component $\sigma$ on site $i$. However, in Bose-Hubbard model simulators, this condition is only fulfilled when the onsite-interaction is sufficiently large and occupations are low, such that the parity of particle number, $\sum_n \ket{2n}\bra{2n}$ is almost equal to the particle number. 

Keeping these limitations in mind, the protocol proceeds as the one for spin systems: Evolving the initial state for time $t_1$, measuring $\hat n_i$, evolving again for time $(t_2-t_1)$ and finally measuring $\hat n_j$, we get (see App.~\ref{app:F_hubb_details} for details) \begin{align}
&\braket{\lbrace \hat n_i(t_1),\hat n_j(t_2)\rbrace}=\notag\\&\braket{\hat n_j(t_2)}_{\ket{1}}\braket{\hat n_i(t_1)}-\braket{\hat n_j(t_2)}_{\ket{0}}(1-\braket{\hat n_i(t_1)})+\braket{\hat n_j(t_2)},
\label{eq:F_nn}
\end{align}
where with $\braket{\hat n_j(t_2)}_{\ket{1}}$ we denoted the expectation value of $\hat n_j$ at time $t_2$ conditioned on having measured occupation one at time $t_1$. The last term is the expectation value of $\hat n_j$ at time $t_2$ without having measured at time $t_1$. 

Non-destructive local projective measurements may be executed in quantum gas microscopes by using fluorescence imaging~\cite{norcia_seconds-scale_2019,norcia_microscopic_2018}. Moreover, optical tweezers could be employed as we detail in App.~\ref{app:F_hubb_details}. Finally, bilayer microscopy~\cite{PhysRevA.91.041602,hartke_measuring_2020,koepsell_robust_2020}  might enable such measurements in a spinful Hubbard model. There, the dynamics can be effectively stopped at time $t_1$ by splitting the spin up/down components from each other and simultaneously increasing the lattice depth. After a fluorescence measurement of one of the components (without measuring the other, which can be done by selecting the layer with the focus of the microscope~\cite{PhysRevA.91.041602}), the two layers are reunited to resume the dynamics before splitting them again to measure at a second time $t_2$. This way, a measurement of $\sum_{i,j} \langle \lbrace \hat n_{i\sigma}(t_1),\hat n_{j\sigma'}(t_2)\rbrace\rangle$ with $\sigma,\sigma'\in \lbrace\uparrow, \downarrow\rbrace$ can be made.

Similarly to the spin protocol, the measurement at time $t_1$ can be deferred until time $t_2$ by mapping the occupation of a site to a tweezer or a different layer of the optical lattice (see App.~\ref{app:F_hubb_details}) and subsequently measuring whether or not an atom was present at time $t_1$ by measuring the tweezer's occupation at time $t_2$.
\subsubsection*{Randomized measurement protocol for $F$}
The protocol employing randomized measurements presented for spin systems can be applied to Hubbard simulators without any adapations, where the necessary implementation of disorder has been demonstrated in both quantum gas microscopes~\cite{choi_exploring_2016,rispoli_quantum_2019, lukin_probing_2019} and superconducting qubits~\cite{roushan_spectroscopic_2017, chiaro_growth_2019}.

\section{Observing the emergence of fluctuation-dissipation relations\label{sct:appli}}

After having introduced measurement protocols for $F$ and $\rho$, we now show that FDRs can be used to characterize thermalization in current quantum simulation platforms. We test the emergence of the FDR in Eq.~\eqref{eq:FDR} by defining the function
\begin{equation}
\mathrm{FDR}(T,\omega)=\log\left(\frac{1}{F(T,\omega)/\rho(T,\omega)-1/2}+1\right),
\label{eq:FDR_def}
\end{equation}
where $\rho(T,\omega)=\int dt e^{i\omega t} \rho(T+t/2,T-t/2)$ is the two-time spectral function at central time $T=(t_1+t_2)/2$. The FDR demands that $\mathrm{FDR}(T,\omega)=\beta \omega$ in equilibrium with the inverse temperature $\beta$ set by the energy of the initial state
\begin{equation}
\braket{\psi_0|\hat H|\psi_0}=\Tr\left( \frac{\exp(-\beta\hat H)}{Z}\hat H\right).
\label{eq:beta}
\end{equation}
All numerical results have been obtained using exact diagonalization, see App.~\ref{app_ED} for our algorithms to efficiently evaluate two-time functions.

\subsection{Thermalization in the Bose Hubbard model}

One of the first demonstrations of the relaxation of equal-time observables towards their equilibrium expectation values was given in an experiment simulating the Bose-Hubbard model~\cite{trotzky_probing_2012}, hence effectively probing the diagonal part of ETH. Here we study the fluctuation-dissipation relations and hence test the validity of the off-diagonal part of ETH.
\begin{figure}
	\includegraphics{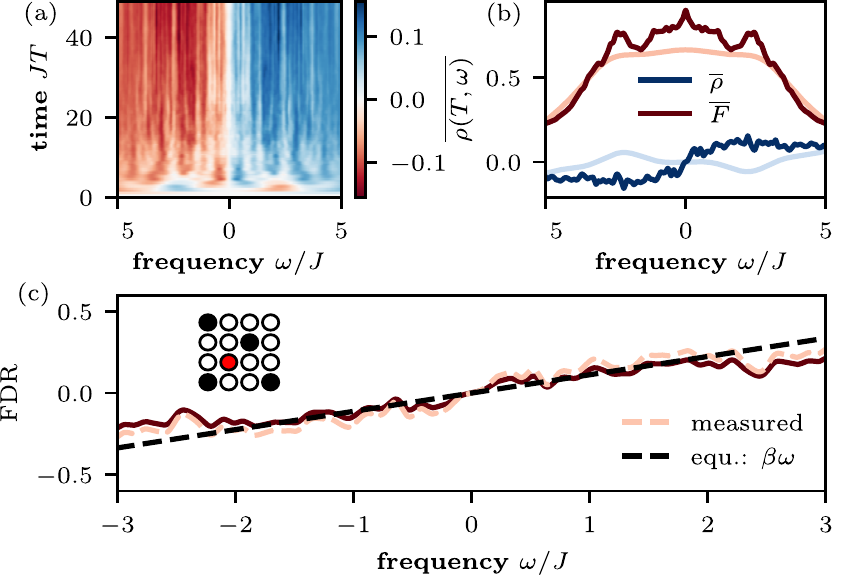}
	\caption{\textbf{Emergence of FDRs in the 2D Bose Hubbard model.} (a) Central-time averaged equal-site density-density spectral function $\rho$ as a function of central time $JT$ and frequency. (b) Late time spectral and statistical functions ($JT=40$, dark) compared to early times ($JT=2$, bright). (c) Fluctuation dissipation relation function defined in Eq.~\eqref{eq:FDR_def} at time $JT=40$ compared to the equilibrium expectation (dashed black line), with the inverse temperature $\beta$ set by energy of the initial state according to Eq.~\eqref{eq:beta}. While the dark red line shows the ideal result, the bright dashed line is the result measured by the linear response and non-destructive projective measurement schemes.  The inset shows the location of the initially occupied sites (black) and the probed lattice site (red) on the $4\times 4$ lattice. The on-site repulsion is given by $U/J=6$. We used a Gaussian frequency broadening with standard deviation $\sigma_\omega=0.05J$ for the Fourier transform. \label{fig:BHM}}
\end{figure}

We study a two-dimensional Bose-Hubbard model with open boundary conditions, given by Hamiltonian
\begin{equation}
\hat H= -J\sum_{<i,j>} (\hat a^\dagger_i \hat a_j+\hat a^\dagger_j \hat a_i) + \frac{U}{2}\sum_{i} \hat n_i(n_i-1),
\end{equation}
where $[\hat a_i,\hat a_j^\dagger]=\delta_{ij}$, $\hat n_i=\hat a^\dagger_i \hat a_i$ and we truncate the local Hilbert space dimension to three states. In Fig.~\ref{fig:BHM} we show the central time averaged statistical and spectral function defined as $\overline{\rho(T,\omega)}=\frac{1}{T}\int_0^T dt \rho(t,\omega)$ for the local density, i.e. $\hat A= \hat B=\hat n_{i}$ with the probed lattice site $i$ indicated in red in Fig.~\ref{fig:BHM}c). In Fig.~\ref{fig:BHM}a) we show that $\overline{\rho}$ (and equally $\overline{F}$, not shown) becomes approximately independent of central time for $JT\gtrsim 20$, indicating that a steady-state has been reached. In order to test whether this steady-state displays the correct connection between $F$ and $\rho$ expected in equilibrium, we plot the FDR function, Eq.~\eqref{eq:FDR_def}, showing that indeed $\mathrm{FDR}(T,\omega)\sim \beta\omega$. The inverse temperature $\beta$ extracted from the FDR matches the expectation from the energy of the initial state (c.f. Eq.~\ref{eq:beta}), indicating that the correct equilibrium state has been reached. Moreover, in Fig.~\ref{fig:BHM}c) we display the FDR function as obtained from an experiment employing non-equilibrium linear response to measure the density-density spectral function $\rho$ and the projective measurement protocol to measure the parity-parity statistical function $F$, which agrees reasonably well with the temperature obtained in the FDR from the ideal case and we find better agreement as the on-site repulsion $U$ is increased.

Here, we showed that full thermalization (i.e. both the diagonal and off-diagonal parts of ETH) can be observed in Hubbard models by probing the emergence of FDRs between the density-density fluctuations and dissipations. In the following, we will discuss cases in which more intricate transient dynamics not contained in the ETH can be observed and characterized via two-time correlation functions.

\subsection{Prethermalization in the long-range transverse field Ising model}
While ETH provides a universal mechanism for how quantum systems reach a thermal steady state at long times, long-lived transient non-thermal states not described by ETH can arise in the dynamics due to a competition of different terms in the Hamiltonian or the presence of non-thermal eigenstates. Here, we will discuss how two-time functions and the FDR can be used to characterize several examples of such \emph{prethermal} steady-states in the long-range transverse field Ising chain (LTFI) implemented in trapped ion quantum simulators 
\begin{equation}
\hat H=\sum_{i<j} \frac{J}{|i-j|^\alpha}\hat \sigma^x_i\hat \sigma^x_j +\frac{g}{2}\sum_i \hat \sigma^z_i
\label{eq:LTFI}
\end{equation}
with chain length $L$, long-range exponent $\alpha$ and transverse field strength $g$. We will discuss how three generic examples of prethermalization can be observed in the FDR, using the LTFI to demonstrate the principle. In the first case, a large transverse field $g$ leads to the classic version of prethermalization as introduced by Berges et al.~\cite{berges_prethermalization_2004}, where a single quasi-conserved quantity prevents full thermalization up to exponentially long times in $J/g$~\cite{abanin_rigorous_2017} and prethermalization to an effective Hamiltonian can be observed in the FDR. In the second example, we show that the generalization of this phenomenon to an \emph{extensive} number of approximately conserved quantities in an integrable sector of the LTFI~\cite{PhysRevE.95.052116} can be used to experimentally diagonalize the Hamiltonian. In the third case, we discuss quenches from a polarized state at $g=0$ to small $g$ and show how emergent confined excitations can be identified by genuine non-equilibrium features in the two-time functions and by a violation of the FDR up to long times.

\begin{figure}
	\includegraphics{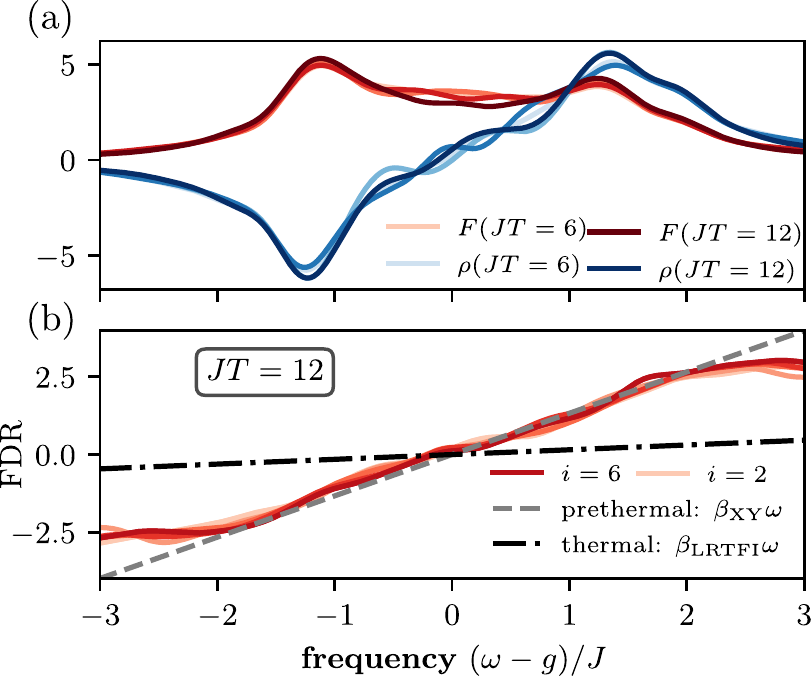}
	\caption{\textbf{Prethermal FDRs in the long-range transverse field Ising model.} (a) The equal-site spectral and statistical functions $\rho, F$ of the local spin raising operators $\hat \sigma^+_i$ with initial state $|\psi_0\rangle=|\uparrow\downarrow\cdots\downarrow\uparrow\rangle_x$ and $i=2$. Central times $JT$ increase from bright to dark. (b) Fluctuation-dissipation relation for different lattice sites $i$ (increasing from bright to dark) along with the expected inverse temperatures $\beta$ in thermal equilibrium of the LTFI and the prethermal Hamiltonian (XY model). We used a Gaussian broadening with standard deviation $\sigma_\omega=0.2J$ for the Fourier transform. Parameters used are  $L=13$ (open boundary conditions), long-range exponent $\alpha=1.5$, transverse field $g=12J$. \label{fig:LRTFI_prethermal_FDR}}
\end{figure}

\subsubsection{Prethermalization due to an approximate conservation law}
Here we study the LTFI in the regime of large transverse field, $g=12J$, and choose the local spin raising/lowering operators $\hat A=\hat \sigma^+_i=\hat B^\dagger$ as operators in the two-time functions, with $\hat\sigma^\pm=\frac{1}{2}(\hat\sigma^x\pm i\hat\sigma^y)$. In Fig.~\ref{fig:LRTFI_prethermal_FDR}a) we show $F$ and $\rho$ starting from the initial state $|\psi_0\rangle=|\uparrow\downarrow\cdots\downarrow\uparrow\rangle_x$, showing that for central times as small as $JT=6$ a steady state has been reached. However, contrary to the case in the Bose-Hubbard model, they are not centred around $\omega=0$. Moreover, the FDR function, shown in Fig.~\ref{fig:LRTFI_prethermal_FDR}b), approximately shows the linear-in-frequency behavior expected  in equilibrium, but with the inverse temperature $\beta$ not matching the expectation from inserting the LTFI into Eq.~\eqref{eq:beta}. Both of these features are explained by the phenomenon of prethermalization~\cite{berges_prethermalization_2004}. Here, the large value of $g$ energetically disfavors all terms in the Hamiltonian changing the total transverse magnetization $\hat S^z=\frac{1}{2}\sum_i \hat \sigma^z_i$, i.e. terms $\sim \sigma^+\sigma^+, \sigma^-\sigma^-$. This leads to an almost conservation of the transverse magnetization and the system effectively evolves under the Hamiltonian $\hat H_{\mathrm{eff}}=\hat H_{XY}+g\hat S^z$ with $\hat H_{\mathrm{XY}}=\sum_{i<j} \frac{J}{|i-j|^\alpha}(\hat \sigma^+_i\hat \sigma^-_j+h.c.)$. The shift of the frequency-space two-time functions follows from the fact that $\hat\sigma^\pm$ are the raising/lowering operators corresponding to the approximate conservation law, i.e. $[\hat S^z,\hat\sigma^\pm]=\pm \hat\sigma^\pm$. Using that $[\hat H_{\mathrm{XY}},\hat S^z]=0$, we find that the term $\sim S^z$ in $H_\mathrm{eff}$ then leads to a precession of the two time functions of $\sigma^{\pm}$, i.e. $\hat \sigma^+(t_1) \hat\sigma^-(t_2) =e^{ig(t_1-t_2)} \hat \sigma'^{+}(t_1) \hat\sigma'^{-}(t_2)$, with the $'$ indicating the remaining non-trivial time evolution with $\hat H_{\mathrm{XY}}$. After the Fourier transform with respect to $t_1-t_2$, this precession leads to the shift $\omega\rightarrow \omega+g$ in the two-time functions and is a direct consequence of the approximate conservation law~\footnote{In general, one would also expect such a shift if the (pre-)thermal steady state has non-zero chemical potential. In our case, however, the chemical potential corresponding to the initial state vanishes such that the shift is solely due to the precession induced by term $~\hat S^z$ in the Hamiltonian.}. From this picture, we moreover expect the system to thermalize to a grand-canonical equilibrium state $e^{-\beta (\hat H_\mathrm{XY}-\mu \hat S^z)}$ instead of $e^{-\beta \hat H}$ on short timescales, where $\mu=0$ for our initial state. This behavior is directly reflected in the temperature found in the FDR: The temperature obtained from inserting $\hat H_\mathrm{XY}$ into Eq.~\eqref{eq:beta} agrees well with the time evolved quasi steady-state (grey dashed line in Fig.~\ref{fig:LRTFI_prethermal_FDR}). We note that at exponentially long times in $J/g$, prethermalization to $e^{-\beta \hat H_\mathrm{XY}}$ would ultimately give way to full thermalization to the LTFI, however we did not find this for our finite-size system on the studied timescales.

In this section, we have shown that the presence of a prethermal conserved quantity can be observed by measuring the FDR corresponding to the raising/lowering operators of the conserved quantity. In the following, we show that this scheme can be generalized to the case of an extensive number of conserved quantities in an integrable model.

\subsubsection{Prethermalization in the vicinity of integrability: Generalized Gibbs ensemble FDR} 

Integrable models possess an extensive (and complete) set of local conserved quantities $\hat{\mathcal{I}}_q$, which prevents them from thermalizing in the sense of the ETH~\cite{rigol_thermalization_2008}. However, integrable models are still expected to fulfill Jayne's maximum entropy principle~\cite{PhysRev.106.620} and hence be described by a ``generalized Gibbs ensemble'' 
\begin{equation}
\hat\rho_{GGE}\sim \exp\left(-\sum_q \lambda_q \hat{\mathcal{I}}_q\right)\end{equation}
at late times with the Lagrange multipliers $\lambda_q$ determined by the initial condition according to $\braket{\psi_0|\hat{\mathcal{I}}_q|\psi_0}\stackrel{!}{=}\Tr(\hat\rho_{\mathrm{GGE}}\hat{\mathcal{I}}_q)$. 

It was shown~\cite{essler_dynamical_2012,PhysRevE.95.052116} that the reasoning for deriving the FDR in App.~\ref{app:KMS_FDR} while replacing the canonical density matrix $\frac{1}{Z}e^{-\beta \hat H}$ with $\rho_{GGE}$ leads to a ``generalized Gibbs ensemble FDR''
\begin{equation}
\tilde F(t_1,t_2)= n_{\lambda_q} \rho(t_1,t_2), \,\, \mathrm{with}\,\, n_{\lambda_q}=\frac{1}{2}+\frac{1}{e^{-\lambda_q}-1}
\end{equation}
for the raising/lowering operators $\hat A=\hat d_q=\hat B^\dagger$ corresponding to the conserved quantities $\hat{\mathcal{I}}_q=\hat d^\dagger_q\hat d_q$ and we defined $\tilde F=\frac{1}{2}\braket{\lbrace \hat d_q(t_1), \hat d_q^\dagger(t_2)\rbrace}$, i.e. $F$ in Eq.~\eqref{eq:F} without subtracting the disconnected part.

For a non-interacting model of the form
\begin{equation}
\hat H= \sum_q \epsilon_q  \hat d_q^\dagger \hat d_q,
\end{equation}
the spectral and statistical functions for $\hat A=\hat d_q=\hat B^\dagger$ trivially fulfill the GGE FDR for all times and initial states. We will show in the following that the GGE FDR is observable in an integrable sector of the LTFI~\footnote{In Ref.~\cite{PhysRevE.95.052116,de_nardis_probing_2017} a similar reasoning was followed, however focussing on density-density (two-particle) correlations from which only linear combinations of the $\lambda_q$ can be extracted. Here, we show instead that the $\lambda_q$ can be directly extracted from the single-particle two-time correlation functions while additionally enabling the extraction of the Bogoliubov angles.}.
\begin{figure*}
	\includegraphics{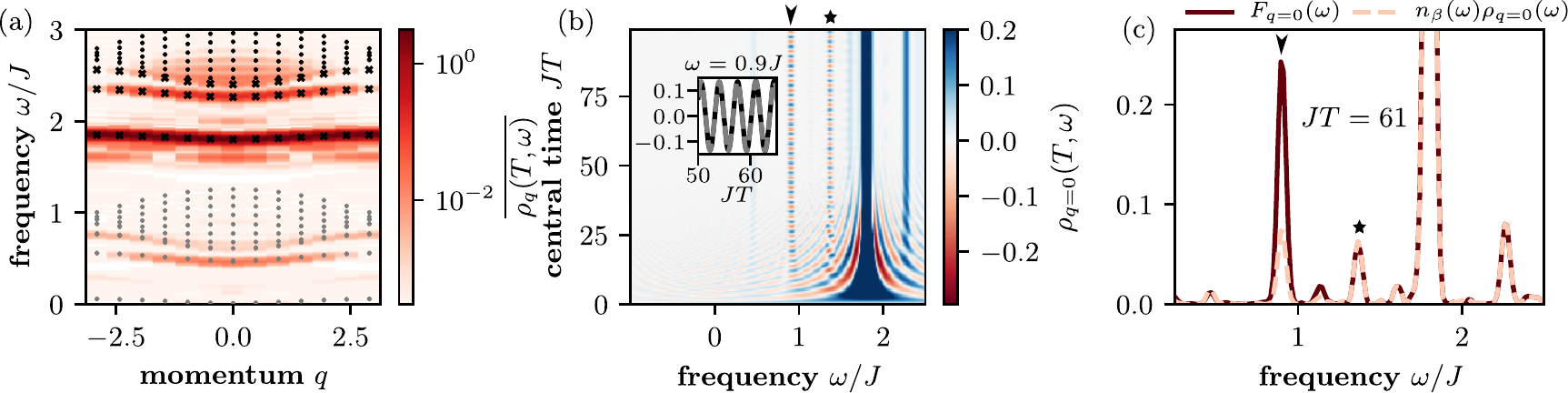}
	\caption{\textbf{Violation of the FDR due to confined excitations in the LTFI.} (a) Two-domain wall spectral function averaged over central time $T$ starting from a completely x-polarized state. Black dots and crosses indicate the difference $E_n-E_0$ of the eigenenergies $E_n$ with the ground state energy $E_0$, grey dots are $E_n-E_1$. (b) Time and frequency resolved spectral function. Black arrow and star indicate oscillatory features, which are non-thermal as depend on central time. Inset: The oscillation frequency of the peak at $\omega=0.9J$ (black) matches the energy difference between the first and second excited state (sinusoidal fit with fixed frequency in dashed grey). (c) Test of the FDR by comparing $F_{q=0}(T,\omega)$ with the corresponding right hand side of the FDR in Eq.~\eqref{eq:FDR} for a fixed time $JT=61$. The black arrow/star indicates the location of the two non-thermal features in b). For all subplots, we used a Gaussian broadening with standard deviation $40/J$ in time and do not plot very small frequencies in c) due to artefacts from the Fourier transform. Here, the transverse field is $g=0.53J$ and the long range exponent $\alpha=2.3$. \label{fig:conf}}
\end{figure*}

For the completely z-polarized state with just a single spin flip, the LTFI is integrable for large fields $g$~\cite{jurcevic_spectroscopy_2015,neyenhuis_observation_2017} and can be solved by employing Holstein Primakoff bosons, $\hat \sigma^z_i\rightarrow 2\hat a_i^\dagger \hat a_i -1$, $\hat \sigma^+_i\rightarrow\hat a_i^\dagger.$ The Hamiltonian can then be diagonalized in momentum space with a Bogoliubov rotation to operators $\hat d_q$ defined by $\hat a_q=\cosh(\Theta_q)\hat d_q -\sinh(\Theta_q)\hat d_q^\dagger$. This results in $\epsilon_q=\sqrt{g(g+2\nu_q)}$, with $\nu_q$ the eigenvalues of $J_{ij}$. Experimentally, only correlation functions corresponding to the unrotated operators $\hat a_q=\hat{\sigma}^-_q$ can be accessed. Solving the non-equilibrium dynamics exactly (see App.~\ref{app:GGE}), we find the spectral function for $\hat A=\hat{\sigma}^-_q$ to be independent of both time and initial state,
$\rho(\omega)=2\pi\cosh^2(\Theta_q)\delta(\omega-\epsilon_q)-2\pi\sinh^2(\Theta_q)\delta(\omega+\epsilon_q)$. Moreover, we find 
$\tilde F(\omega,T\rightarrow\infty)=\left(2\pi\cosh^2(\Theta_q)\delta(\omega-\epsilon_q)+2\pi\sinh^2(\Theta_q)\delta(\omega+\epsilon_q)\right)(\frac{1}{2}+\braket{\psi_0|\hat d^\dagger_q\hat d_q|\psi_0})$ for the statistical function at large central times. Hence, even these unrotated, experimentally accessible spectral and statistical functions fulfill the GGE FDR as $n_{\lambda_q}=1/2+\braket{\psi_0|\hat d^\dagger_q\hat d_q|\psi_0}$ by definition of the GGE. Moreover, the Hamiltonian can be ``experimentally diagonalized'' by measuring $F$ or $\rho$ as the dispersion $\epsilon_q$ can be read off from the position of the peaks and the Bogoliubov angles $\Theta_q$ from the ratio of the height of the peaks at positive and negative frequency~\footnote{Note that to read off $\Theta_q$, only one-time correlation functions need to be measured as $\rho(\omega)$ is independent of the central time $T$.}.

Note that ``thermalization'' to a GGE is in practice only a transient phenomenon as there are always integrability-breaking terms present in experiment~\cite{kinoshita_quantum_2006,gring_relaxation_2012}, leading to thermalization to a (grand-)canonical ensemble at late times. This is why the case discussed here is a direct generalization of the prethermalization discussed in the previous section, with the single additional conservation law replaced with extensively many.

So far, we showed that the nature of the (pre-)thermal steady state can be elucidated from measuring FDRs, hence showing their potential to test the assumptions of the ETH. In the following, we will show that information contained in $F$ and $\rho$ can also be used to identify the relevant excitations for the thermalization dynamics in a case in which violations of the FDR (and therefore the ETH) survive up to long times.

\subsubsection{Prethermalization due to confined excitations} At small transverse fields, $g<J$, the LTFI shows confinement of domain wall excitations ~\cite{kormos_real-time_2017}, which leads to non-thermal eigenstates in the spectrum~\cite{james_nonthermal_2019} and and long thermalization times~\cite{mazza_suppression_2019,liu_confined_2019,lerose_quasilocalized_2019}. Here, we will show that this effect reminiscent of the confinement between quarks in QCD~\cite{PhysRevD.10.2445} leads to non-thermal features in two-time correlation functions, including a violation of the FDR (and hence ETH) up to long times. The proposal discussed here for characterizing confined excitations by two-time correlations may be used in the future to characterize unknown non-thermal eigenstates directly in experiment.

We prepare the totally x-polarized initial state $\ket{\psi_0}=\ket{\uparrow \cdots \uparrow}_x$, which is close to one of the two ground states due to $g<J$. We directly probe the confined excitations by calculating the two-domain-wall spectral and statistical function in momentum space by choosing $\hat A\equiv \hat \sigma_2^+= (\hat\sigma^z_q+i\hat \sigma^y_q)/2=\hat B^\dagger$, which flips a spin and hence creates two domain walls. In Fig.~\ref{fig:conf}a) we show the central-time averaged non-equilibrium spectral function for $\alpha=2.3$, $g=0.53J$ and periodic boundary conditions (replacing the distance $|i-j|$ in Eq.~\eqref{eq:LTFI} with $\mathrm{min}(|i-j|,L-|i-j|)$). Three nearly dispersionless sharp excitations (linewidth limited by the numerical broadening) between $\omega\approx 1.9J$ and $\omega\approx 2.3J$ are clearly visible along with a continuum of excitations above them. These correspond to excitations within and outside of the confining potential, respectively 
~\cite{liu_confined_2019}, as we show by plotting the difference between the excited state eigenenergies and the ground state energy $E_n-E_0$ (black crosses for confined excitations, black dots for continuum). Moreover, we find some spectral weight \emph{below} the gap ($\omega\approx 1.9J$), at frequencies corresponding to the energy difference of the eigenenergies with the first excited state $E_n-E_1$. Moreover, we find oscillations of the spectral weight as a function of central time in Fig.~\ref{fig:conf}b) (marked by an arrow and a star), indicating that an equilibrium state has not yet been reached up to times as along as $JT=100$ ~\footnote{Oscillations for times JT<30 around the peak at $\omega/J=1.9$ are artifacts of the Fourier transform (Gibbs phenomenon).}. This is further substantiated by a violation of the FDR at the location of some of these oscillatory features (Fig.~\ref{fig:conf}c)). In the following, we will show that these non-thermal features are a direct consequence of the large overlap of the initial state with sharp excitations and show that their properties can be read off from the two-domain-wall nonequilibrium spectral function. First, note that the Lehmann representation of the spectral function can be split into a time dependent and time independent part~\cite{PhysRevB.98.224205}, resulting in
\begin{align}
\rho(T,\omega)&=\sum_n |\langle \psi_0|n\rangle|^2 \rho_{nn}(\omega)\notag \\&+\sum_{n,m\neq n} \langle \psi_0|n\rangle \langle m|\psi_0\rangle e^{i(E_n-E_m)T}\rho_{nm}(\omega)
\label{eq:nonequ_specfunc}
\end{align}
with the eigenstate spectral functions
\begin{align}
\rho_{nm}(\omega) =\sum_l &\langle n| \hat \sigma_2^+|l\rangle\langle l|\hat \sigma_2^-|m\rangle \delta(\omega-(E_l-\frac{E_m+E_n}{2}))\notag\\-&\langle n| \hat \sigma_2^-|l\rangle\langle l| \hat \sigma_2^+|m\rangle\delta(\omega+(E_l-\frac{E_m+E_n}{2})).
\label{eq:eigenstatespec}
\end{align} 
From the time independent/diagonal part we can directly explain the spectral weight below the gap: Because of the large overlap of the initial state with the first excited state $\ket{1}$, also $\rho_{11}(\omega)$ contributes, which contains delta-peaks at frequencies $E_n-E_1$. 
Furthermore, the only central time-dependence is contained in an oscillatory term with frequency $E_n-E_m$, which appear at frequencies $\omega'$ given by a superposition of \emph{three} eigenergies, $\omega'=\pm (E_l-(E_m+E_n)/2)$. We can use this observation to analyse the oscillatory features found in the non-equilibrium spectral function. At $\omega'\approx 0.9$ (marked by an arrow) and $\omega'\approx 1.4$ (marked by a star) we find that the central time oscillation frequency is in perfect agreement with $E_0-E_1$, indicating that $m,n \in {0,1}$. From the frequency location $\omega'$, we can furthermore identify that $l=0,1$ and $l=3$ are the contributions in Eq.~\eqref{eq:eigenstatespec} leading to the features at $\omega'\approx 0.9$, $\omega'\approx 1.4$, respectively. This shows that the central time oscillations arise solely from the two lowest excited states corresponding to confined excitations~\footnote{Note that the two lowest eigenstates are doubly degenerate, leading to the non-zero matrix elements between states of equal energy needed to give a non-zero contribution in Eq.~\eqref{eq:eigenstatespec}. To illustrate this, consider the two ground states $\ket{0},\ket{0'}$ (given by $\ket{\uparrow\cdots\uparrow}_x$, $\ket{\downarrow\cdots\downarrow}_x$ for $g=0$), $\braket{0|\hat \sigma_2^+|0'}\neq 0$ because $\hat \sigma_2^+$ flips a spin, $\hat \sigma_2^+ \ket{\uparrow}=\ket{\downarrow}$ and the resulting state has non-zero overlap with the other ground-state because of $g\neq 0$.}. In general, one would expect such central-time oscillations to dephase rapidly. Here, however, the fact that the initial state has a strongly peaked overlap with eigenstates well isolated in energy leads to a long lifetime of the central-time oscillations.

While any such central-time dependent contribution leads to a deviation from the diagonal ensemble (which is the first term in Eq.~\eqref{eq:nonequ_specfunc}) and hence a lack of thermalization, the FDR is not necessarily violated if $\rho$ and $F$ are shifted equally (assuming the individual eigenstates fulfill the FDR). Indeed, as visible in Fig.~\ref{fig:conf}c), the oscillatory feature at $\omega\approx 0.9J$ violates the FDR while the one at $\omega\approx 1.4J$ does not, despite having the same oscillation amplitude and frequency. This is explained by comparing the expression in Eq.~\eqref{eq:eigenstatespec} with the corresponding one for $F$, given by \begin{align}
F(T,\omega)&=\sum_n |\langle \psi_0|n\rangle|^2 F_{nn}(\omega)\notag \\&+\sum_{n,m\neq n} \langle \psi_0|n\rangle \langle m|\psi_0\rangle e^{i(E_n-E_m)T}F_{nm}(\omega)
\label{eq:nonequ_specfunc}
\end{align}
with the eigenstate statistical functions
\begin{align}
F_{nm}(\omega) =\frac{1}{2}\sum_l &\langle n| \hat \sigma_2^+|l\rangle\langle l|\hat \sigma_2^-|m\rangle \delta(\omega-(E_l-\frac{E_m+E_n}{2}))\notag\\+&\langle n| \hat \sigma_2^-|l\rangle\langle l| \hat \sigma_2^+|m\rangle\delta(\omega+(E_l-\frac{E_m+E_n}{2})).
\label{eq:eigenstatestat}
\end{align}
The only difference to $\rho$ is an overall factor $1/2$ (which would get compensated on the right-hand-side of the FDR by $n_{\beta}(\omega)  \approx 1/2$ at low temperatures) and the two terms in the first and second line in Eq.~\ref{eq:eigenstatestat} get added instead of subtracted. By explicitly analyzing the contributions in Eq.~\eqref{eq:eigenstatespec}, we found that for the feature at $\omega\approx 1.4J$ the first term dominates, which has the same sign in the expressions for $F$ and $\rho$ such that both get shifted equally compared to the diagonal ensemble expectation and the FDR is fulfilled. Contrastingly, for the feature at $\omega\approx 0.9J$, the second term dominates, which has a different sign in $F$ and $\rho$ such that the FDR is violated. In Fig.~\ref{fig:conf}c), we find a second FDR-violating feature around $\omega\approx 1.1J$, with an oscillation frequency matching $E_2-E_0$, corresponding to contributions from the ground state and second confined state, $n,m,l \in 0,2$. Note that the violation of the FDR we observe here can not be explained by an effective, i.e. frequency independent temperature differing from the one expected from the energy of the initial state, which for example occurs in periodically driven systems~\cite{PhysRevLett.124.106401}. Such an effective non-thermal temperature would manifest its-self in a mismatch of $F$ and the right-hand-side of the FDR, $n_\beta(\omega)\rho$, for all frequencies low enough to show the $\beta$ dependence of $n_\beta(\omega)$ (i.e. such that $n_\beta(\omega)$ differs significantly from $1/2$). This is however not the case here: in Fig.~\ref{fig:conf}c) a peak at frequency $\omega\approx0.25J$ \emph{fulfilling} the FDR is clearly visible, showing that the violations of the FDR discussed here indeed occur at isolated frequencies and cannot be explained by a non-thermal effective temperature.

For most of the interpretations given above, no additional numerics apart from the calculation of the two-time functions were needed and the same conclusions could have been made only given an experimental measurement of the two-time functions. Therefore, this provides a general procedure how to extract information about long lived prethermal (or even non-thermal) excitations completely independently of numerical calculations: Central time oscillations indicate their presence while the central time oscillation frequency and frequency location $\omega'$ can be used to extract their energy. The property that the FDR is violated or not at the location of the peak can then be used to extract information about the matrix elements and hence about the nature of the excitation itself, where the latter can be refined by probing two-time correlations of different operators and initial states.

\section {Conclusions and Outlook}
We have shown how to probe the off-diagonal part of eigenstate thermalization with two-time functions in quantum simulators, which is an open experimental challenge. We discussed and introduced measurement protocols in quantum simulators of spin and Hubbard models for the two-time spectral function $\rho$ and statistical function $F$, which are in general independent of each other out of equilibrium. We have shown that probing the link between the statistical function $F$ and the spectral function $\rho$ via the fluctuation-dissipation relations can be used to probe the off-diagonal part of ETH independently of both microscopic details and theory input, thus providing a general route to probing thermalization in quantum simulators. Going beyond testing thermalization of the steady-state at long times, we showed that the FDRs can also be used to characterize prethermal steady states, which can lead to modifications of the FDR in the case of almost conserved quantities and can even lead to a violation of the FDR in the presence of confined excitations.

Our scheme can be used to probe multiple aspects of thermalization. By preparing initial states with energy densities covering the whole spectrum (for example spin spirals~\cite{PhysRevLett.113.147205,babadi_far--equilibrium_2015,brown_two-dimensional_2015, jepsen_spin_2020}, thermalization of a many-body Hamiltonian across its whole energy spectrum could be probed. Individual eigenstates could be prepared by a recently proposed protocol employing weak measurements~\cite{yang_quantum_2020}), thus opening the route to directly test the off-diagonal part of ETH in terms of individual eigenstates with the FDR. In many-body localized systems, a uniform late-time temperature is not expected, however, local temperatures can be defined~\cite{PhysRevLett.121.267603} and could be measured by using the FDRs as a local thermometer. Two-time functions show aging in classical glasses~\cite{PhysRevLett.88.257202,PhysRevX.10.021051}, their measurement could hence probe the analogy to glasses made in quantum systems with slow relaxation~\cite{PhysRevLett.123.040601,signoles_glassy_2020}. Furthermore, the non-thermal oscillatory features we found for confined excitations could be used to characterize other non-thermal states such as many-body scars~\cite{bernien_probing_2017,turner_weak_2018}. Our measurement protocols could also be used to show violations of the FDR due to transport processes near non-thermal fixed points~\cite{PhysRevLett.122.150401}.  Lastly, our protocols offer a route to quantum simulate pump-probe experiments on solids such as optical spectroscopy~\cite{eckstein_theory_2008} (measuring $\rho$)  and optical noise spectroscopy~\cite{randi_probing_2017} (measuring $F$) by using the analogy between the light-matter couplings and the resulting linear-response correlation functions. While in the solid state, the non-zero charge of the electron   leads to a coupling of the current density to the light field, in cold (neutral) atom platforms, the dipolar coupling leads to a coupling of the atom number density to the light. Hence, the measurement of density-density two-time functions proposed here is analogous to the current-current functions of optical measurements in the solid state.

\paragraph*{Acknowledgments.--}We thank Diego Barberena, J\"urgen Berges, Annabelle Bohrdt, Hope Bretscher, Eleanor Crane, Andreas Elben, Johannes Feldmeier, Titus Franz, Martin G\"arttner, Fabian Grusdt, Clemens Kuhlenkamp, Siddharth Parameswaran, Asier Pi\~neiro Orioli, Pablo Sala, Tibor Rakovszky,  Marcos Rigol, Antonio Rubio-Abadal, Philipp Uhrich, Jayadev Vijayan, Aaron Young for insightful discussions. We used the package QuSpin~\cite{SciPostPhys.2.1.003,10.21468/SciPostPhys.7.2.020} for some of our numerics. We acknowledge support from the Max Planck Gesellschaft (MPG) through the International Max Planck Research School for Quantum Science and Technology (IMPRS-QST), the Technical University of Munich - Institute for Advanced Study, funded by the German Excellence Initiative, the European Union FP7 under grant agreement 291763, the Deutsche Forschungsgemeinschaft (DFG, German Research Foundation) under Germany's Excellence Strategy--EXC-2111--390814868, the European Research Council (ERC) under the European Union’s Horizon 2020 research and innovation programme (grant agreement No. 851161), from DFG grant No. KN1254/1-1, No. KN1254/2-1, and DFG TRR80 (Project F8). 

\appendix
\section{The KMS condition and fluctuation-dissipation relations\label{app:KMS_FDR}}
The fluctuation-dissipation relations are a consequence of the cyclicity of the trace and the interpretation of two-time correlators in terms of spectral and statistical components, which follow from the commutation relations.

\paragraph*{Kubo-Martin-Schwinger (KMS) condition in thermal equilibrium.--} The KMS condition for a correlation function of two operators $\hat A(t_1)$ and $\hat B(t_2)$ evaluated in the Heisenberg picture with Hamiltonian $\hat H$ is a simple property of the thermal density matrix:
\begin{align}
\Tr{\left[e^{-\beta \hat H}\hat A(t_1)\hat B(t_2)\right]}&=
\Tr{\left[e^{-\beta \hat H}e^{\beta \hat H}\hat B(t_2)e^{-\beta \hat H}\hat A(t_1)\right]}\notag\\
&=\Tr{\left[e^{-\beta \hat H}\hat B(t_2-i\beta)\hat A(t_1)\right]},
\label{eq:KMS_time}
\end{align}
where we only used the cyclicity of the trace. In particular, the above relation does not depend on the commutation relations of $\hat A$ and $\hat B$ (This is in general not true if above relations are defined in terms of a path integral as then all correlation functions are automatically time ordered and the fermionic relation (i.e. for $\hat A$,$\hat B$ being fermionic creation/annihiliation operators) acquires a minus sign~\cite{berges_nonequilibrium_2015}.)

Defining the two Wightman functions by using time-translational invariance of thermal equilibrium,
\begin{align}
G^>(t_1-t_2)&=\frac{1}{Z}\Tr{\left[e^{-\beta \hat H}\hat A(t_1)\hat B(t_2)\right]}\\
G^<(t_1-t_2)&=\frac{1}{Z}\Tr{\left[e^{-\beta \hat H}\hat B(t_2)\hat A(t_1)\right]},
\end{align}
with $Z=\Tr e^{-\beta \hat H}$, and Fourier transforming with respect to $t_1-t_2$, $G^>(\omega)=\int\mathrm{d}t e^{i\omega t}G^>(t)$, the KMS condition simply becomes
\begin{equation}
G^>(\omega)=e^{\beta\omega}G^<(\omega).
\label{eq:simple_KMS}
\end{equation}

\paragraph*{Fluctuation dissipation relations (FDRs).--} FDRs may be obtained from the KMS condition by combining the Wightman functions into (bosonic or fermionic) spectral ($\rho$) and statistical ($F$) components as
\begin{align}\rho(\omega)&:=G^>(\omega)\mp G^<(\omega)\\
F(\omega)&:=\frac{1}{2}\left(G^>(\omega)\pm G^<(\omega)\right),
\end{align}
where the upper (lower) sign corresponds to bosons (fermions), respectively. These definitions respect the proper interpretation of $\rho$ as a spectral function as may be motivated from the sum rule $\int\frac{d\omega}{2\pi}\rho(\omega)=\rho(t=0)=\langle[\hat A, \hat B]_{\mp}\rangle$, i.e., the equal-time (anti-)commutation relations.

Inserting the KMS condition in Fourier space into above definitions, we find the FDRs
\begin{equation}
F(\omega)=n_\beta(\omega) \rho(\omega),\label{eq:FDR_supp}
\end{equation}
with $n_\beta(\omega)=\frac{1}{2}\pm 1/(\exp(\beta\omega)\mp 1)$ the Bose-Einstein/Fermi-Dirac distribution at inverse temperature $\beta$. We emphasize that whether bosonic or fermionic FDRs are obtained is not a mathematical property of the operators $\hat A$ and $\hat B$ but of the physical interpretation of the (anti-)commutator as the spectral/statistical function. In particular, this interpretation is ambiguous in the case of spin operators due to the sum rules differing between equal-site and un-equal site operators. For example, the raising/lowering operators $\hat \sigma^{\pm}_i$ anticommute for equal sites but commute for un-equal sites. Conventionally, bosonic FDRs are used for spin systems~\cite{babadi_far--equilibrium_2015}, which we also follow here.

We furthermore note that the FDR is not defined at $\omega=0$ as the KMS condition in Eq.~\eqref{eq:simple_KMS} implies $\rho(\omega=0)=0$, with $F(\omega=0)$ left unconstrained. 

\section{FDRs and the eigenstate thermalization hypothesis~\label{app:ETH_FDR}}

Here we summarize the arguments in Ref.~\cite{dalessio_quantum_2016} to show that the eigenstate thermalization hypothesis (ETH) implies the FDRs, and that the experimental test of FDRs directly tests the off-diagonal part of ETH. We supplement the analytical arguments by showing the FDR on the level of individual eigenstates in the two-dimensional Bose Hubbard model.

To prove these statements, we assume $\hat{B}=\hat{A}^\dagger$, which is the case for all functions evaluated in the main text. For general $\hat{B}\neq \hat A$ additional assumptions not contained in the ETH have to be made~\cite{dalessio_quantum_2016}. For late times $T$, all $T$ dependent terms in the Lehmann representation of the spectral and statistical functions are expected to dephase (c.f. Eq.~\eqref{eq:nonequ_specfunc}), such that 
\begin{align}
\lim_{T\rightarrow\infty}F(T,\omega)\equiv &F(\omega)=\sum_n |\braket{\psi_0|n}|^2 F_{nn}(\omega),\label{eq:longtimeF}\\
 \lim_{T\rightarrow\infty}\rho(T,\omega)\equiv &\rho(\omega)=\sum_n |\braket{\psi_0|n}|^2 \rho_{nn}(\omega).
\end{align}
with the eigenstate spectral/statistical functions given by
$F_{nn}(\omega) =\frac{1}{2}\sum_{l\neq n} |\langle n|\hat A|l\rangle|^2(\delta(\omega-(E_l-E_n))+\delta(\omega+(E_l-E_n)))$, $\rho_{nn}(\omega) =\sum_{l\neq n} |\langle n|\hat A|l\rangle|^2( \delta(\omega-(E_l-E_n))-\delta(\omega+(E_l-E_n)))$. This expression makes explicit that the long-time value of the spectral and statistical functions is entirely determined by the off-diagonal matrix elements of $\hat A$. Comparing this with the corresponding equilibrium expressions $F_\mathrm{equ.}(\omega)= \frac{1}{Z}\sum_n e^{-\beta E_n} F_{nn}(\omega)$, $\rho_\mathrm{equ.}(\omega)=\frac{1}{Z}\sum_n e^{-\beta E_n} \rho_{nn}(\omega)$, one may first be lead to believe that the $|c_n(0)|^2$ must correspond to the weights in thermal equilibrium, $\frac{1}{Z}e^{-\beta E_n}$, in order for the equilibrium FDR to hold. This is however in general not true, as the $|c_n(0)|^2$ do not resemble any of the thermal ensembles~\cite{rigol_quantum_2009} for most physical initial states. The eigenstate thermalization hypothesis offers a different route to thermalization in the sense of FDRs: \emph{each eigenstate fulfills an FDR individually} and hence the weighted sum over the initial state distribution $|c_n(0)|^2$ does so, too.

 Now, consider the Fourier transformed correlation function of a single eigenstate,
\begin{align}
C_n(\omega)&=\int \mathrm{d}\tau e^{i\omega \tau} \braket{n|\hat A(\tau)\hat A^\dagger(0)|n}\\
&= 2\pi \sum_m \delta(\omega-(E_m-E_n)) |\braket{n|\hat A|m}|^2.
\label{eq:Cn}
\end{align}
The ETH Ansatz~\cite{srednicki_chaos_1994} demands that
\begin{equation}
\braket{n|\hat A|m} = A(\bar E)\delta_{nm} + e^{-S(\bar{E})/2}f_A(\bar E,E_m-E_n)R_{nm},
\end{equation}
where $A(\bar E)$ is the microcanonical expectation value of operator $\hat A$ at energy $\bar{E}=(E_n+E_m)/2$, $S$ is the thermodynamic entropy, $R_{nm}$ are random numbers with mean zero and unit variance and $f_A(\bar E,E_m-E_n)$ and $A(\bar E)$ are smooth functions of their arguments $\bar E$. Inserting this ansatz into the eigenstate correlation function and replacing the sum over energies by an integral $\sum_m\rightarrow \int d(E_m-E_n) \exp[S(E_n+(E_m-E_n))]$ and using that the $|R_{nm}|^2$ average out under the sum, we then arrive at
\begin{align}
C_n(\omega)/2\pi&= |A(\bar E)|^2\delta(\omega)\notag\\&+ e^{S(E_n+\omega)-S(E_n+\omega/2)} |f_A(E_n+\omega/2,\omega)|^2.
\end{align}
As argued in Ref.~\cite{dalessio_quantum_2016} both $S$ and $f_A$ can be Taylor expanded around $\omega=0$ if $\hat{A}$ is a local few-body operator, such that
\begin{equation}
C_n(\omega)/2\pi= |A(\bar E)|^2\delta(\omega)+ e^{\beta\omega/2}|f_A(E_n,\omega)|^2,
\end{equation}
where we used that $dS(E)/dE=\beta$ with $\beta=\beta(E)$ the inverse temperature.
We construct the eigenstate spectral and statistical functions from $C_n(\omega)$ by using $|f_A(E_n,\omega)|^2|R_{nm}|^2=|f_{A^\dagger}(E_n,-\omega)|^2|R_{mn}|^2$, 
resulting in
\begin{align}
F_{nn}(\omega)/2\pi&= \cosh({\beta\omega/2})|f_A(E_n,\omega)|^2,\\
\rho_{nn}(\omega)/2\pi&= 2\sinh({\beta\omega/2})|f_A(E_n,\omega)|^2.
\end{align}
Both $F$ and $\rho$ are hence entirely determined by $f_A$ and the inverse temperature corresponding to the eigenenergy $E_n$. Moreover, we finally find that the FDR holds on the level of a single eigenstate,
\begin{equation}
F_{nn}(\omega)=n_\beta(\omega)\rho_{nn}(\omega)
\end{equation}
with $n_\beta(\omega)=\frac{1}{2}+1/(\exp(\beta\omega)-1)$.

From this result we can now deduce the conditions on the initial state for the FDR. Inserting the eigenstate FDR into the long-time limit of the non-equilibrium statistical function (c.f. Eq.~\eqref{eq:longtimeF}),
\begin{align}
F(\omega)&=\sum_n |c_n|^2 n_{\beta(E_n)}(\omega)\rho_{nn}(\omega) \\&\stackrel{?}{=}n_\beta(\omega)\rho(\omega),
\end{align}
we clearly see that the second equality can only be true if the $|c_n|^2$ are concentrated around a region in which $\beta{(E_n)}$ is not a strongly varying function.
\begin{figure}
\includegraphics{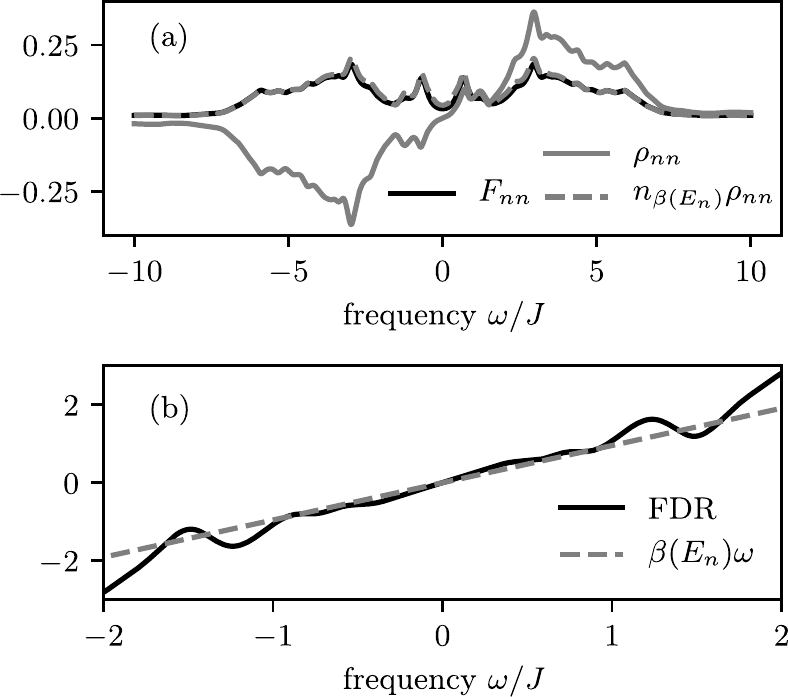}
\caption{\textbf{FDR for a single eigenstate.--} (a) Eigenstate statistical $F_{nn}$ and spectral functions $\rho_{nn}$ as well as the right-hand-side of the FDR $n_\beta\rho_{nn}$ for an eigenstate $n$ with eigenenergy $E_n\approx-9.06J$ in the 2D Bose Hubbard model at $U/J=4$ with the same initial state and operator proped as in Fig.~\ref{fig:BHM} in the main text. The corresponding  inverse temperature $\beta(E_n)\approx 0.95 J$ expected in thermal equilibrium is set by the eigenenergy of the state via Eq.~\eqref{eq:beta}. To evaluate Eq.~\eqref{eq:Cn} we used a Lorentzian broadening with FWHM of $0.2J$. (b) FDR function as defined in Eq.~\eqref{eq:FDR_def} for low frequencies, showing the expected linear behaviour with a slope matching the inverse temperature. \label{fig:Eigenstate_FDR}}
\end{figure}

\paragraph*{Numerical verification of the ETH scenario.--} In Fig.~\ref{fig:Eigenstate_FDR} we verify the FDR for a single eigenstate of the 2D Bose Hubbard model. The inverse temperature $\beta$ extracted from the FDR matches the expectation from the corresponding eigenenergy, i.e. from solving $E_n=\frac{1}{Z} \Tr[ e^{-\beta\hat H}\hat H]$ for $\beta$. See Ref.~\cite{noh_numerical_2020} for an in-depth analysis of finite size effects in the FDR from the perspective of ETH.

\section{Generalized Gibbs Ensemble FDR near the fully polarized state in the LTFI\label{app:GGE}}
Here we show how a generalized FDR can be observed in integrable models, where thermalization to a generalized Gibbs ensemble 
\begin{equation}
\hat\rho_{GGE}=\frac{1}{Z} \exp\left(-\sum_k \lambda_k \hat{\mathcal{I}}_k\right)
\label{eq:GGE_app}\end{equation}
with $Z= \Tr\left[\exp\left(-\sum_k \lambda_k \hat{\mathcal{I}}_k\right)\right]$ is expected. We first prove the general statements made in the main text and then use an integrable limit of the long-range transverse field Ising model (LTFI), $\hat H=\frac{1}{2}\sum_{i\neq j}J_{ij}\sigma^x_i\sigma^x_j+\frac{g}{2}\sum_j \sigma^z_j$ with $J_{ij}=J/|i-j|^\alpha$, as an example to show that observation of the GGE FDR in experiment is possible.

\paragraph*{Generalized KMS condition and FDRs.--} Thermalization to the GGE implies that two-time correlation functions of operators $\hat A$ and $\hat B$ fulfill a generalized KMS condition
\begin{equation}
\Tr\left(\hat\rho_{\mathrm{GGE}} \hat A(t_1) \hat B(t_2) \right)=\Tr\left(\hat\rho_{\mathrm{GGE}}\hat B'(t_2)\hat A(t_1)  \right),
\end{equation}
with $\hat  B'(t_2)=e^{\sum_k \lambda_k \hat{\mathcal{I}}_k}\hat B(t_2) e^{-\sum_k \lambda_k \hat{\mathcal{I}}_k}$.
The resulting FDR then crucially depends on the operator $\hat B$. For example, for $\hat B=\hat{\mathcal{I}}_k$ it follows that $\hat{\mathcal{I}}_k'(t_2)= \hat{\mathcal{I}}_k(t_2)$ and therefore the commutator vanishes, $\rho=\braket{[\hat A(t_1), \hat{\mathcal{I}}_k(t_2)]}_\mathrm{GGE}=0$, rendering the FDR meaningless as the anticommutator $\tilde F=\frac{1}{2}\braket{\lbrace\hat A(t_1), \hat{\mathcal{I}}_k(t_2)\rbrace}_\mathrm{GGE}$ is in general non-zero.

An FDR of the expected form is however obtained for $\hat B= \hat d_k$, defined by $\hat{\mathcal{I}}_k= \hat d^\dagger_k \hat d_k$ as then $\hat d_k'(t_2)=e^{-\lambda_k} \hat d_k(t_2)$ and hence $\braket{\hat A(t_1)\hat d_k(t_2)}=e^{-\lambda_k}\braket{\hat d_k(t_2)\hat A(t_1)}$. Therefore we find the FDR
\begin{equation}
\tilde F(t_1,t_2)= \left(\frac{1}{2}+n_{\lambda_k}\right) \rho(t_1,t_2), \quad \mathrm{with}\quad n_{\lambda_k}=\frac{1}{e^{-\lambda_k}-1}.
\label{eq: FDR_dk}
\end{equation}

\paragraph*{Integrable limit of the LTFI.--} If the initial state has only a few $n$ on top of the fully polarized state in the direction of the field, $\ket{\Psi_0}=\ket{\uparrow\uparrow\cdots\uparrow}$ or $\ket{\Psi_0}=\ket{\downarrow\downarrow\cdots\downarrow}$, the dynamics can be accurately described within linear spin-wave theory~\cite{hauke_spread_2013,jurcevic_quasiparticle_2014,jurcevic_spectroscopy_2015,neyenhuis_observation_2017}. Employing a Holstein-Primakoff transformation $\hat \sigma^z_i\rightarrow 2\hat a_i^\dagger \hat a_i -1$, $\hat \sigma^+_i\rightarrow\hat a_i^\dagger\sqrt{1-\hat a_i^\dagger\hat a_i} \approx \hat a_i^\dagger$, we can map the LTFI Hamiltonian of length $L$ to
\begin{equation}
\hat H = \sum_{i,j} J_{ij} \left(\hat a_i^\dagger\hat a_j +\frac{1}{2}\left(\hat a_i^\dagger\hat a_j^\dagger+\hat a_i\hat a_j\right)\right)+g \sum_i \hat a_i^\dagger\hat a_i
\end{equation} in the regime where $\mathrm{max}(J_{ij})\ll g$ at low filling such that the pairing terms are suppressed and hence the spin-wave approximation stays valid in the dynamics. For a single spin flip on top of the fully polarized state, this mapping becomes exact as $\mathrm{max}(J_{ij})/g\rightarrow 0$.

We diagonalize the spatial degree of freedom by employing an orthogonal transformation $U U^T=\mathbb{1}$, such that $\sum_{i,j}=U_{ik}J_{ij}U_{jk'}=\nu_k \delta_{kk'}$, which introduces a conjugate coordinate $k$ via $\hat a_k=\sum_i U_{ik} \hat a_i$ and $\nu_k$ are the eigenvalues of the interaction matrix $J_{ij}$ ($J_{ii}=0$)~\cite{neyenhuis_observation_2017}. The Hamiltonian then reads
\begin{equation}
\hat H = \sum_k (\nu_k+g) \hat a^\dagger_k \hat a_k + \frac{1}{2} \nu_k(\hat a^\dagger_k \hat a^\dagger_k +\hat a_k \hat a_k )
\label{eq:H_SW_fourier}
\end{equation}
and can be diagonalized via a Bogoliubov transformation $\hat a_k=\cosh(\Theta_k)\hat d_k -\sinh(\Theta_k)\hat d_k^\dagger$, $\Theta_k= \frac{1}{2} \mathrm{arctanh}\left(\nu_k/(\nu_k+g)\right)$ such that
\begin{equation}
\hat H = \sum_k \epsilon_k \hat d_k^\dagger \hat d_k\quad \mathrm{with}\quad \epsilon_k=\sqrt{g(g+2\nu_k)}.
\label{eq:Ham_SW_diag}
\end{equation}

The explicitly diagonalized Hamiltonian in Eq.~\eqref{eq:Ham_SW_diag} shows that the LTFI has extensively many conserved quantities $\hat {\mathcal{I}}_k = \hat n_k \equiv \hat d_k^\dagger \hat d_k$ in this regime, implying that the equilibrium state is described by a GGE (c.f. Eq.~\eqref{eq:GGE_app}).
The Lagrange multipliers $\lambda_k$ to which an initial state $\ket{\Psi_0}$ is expected to thermalize to are determined by the condition $\braket{\Psi_0|\hat n_k|\Psi_0}\equiv \braket{\hat n_k}_0\stackrel{!}{=}\frac{1}{Z}\Tr\left[\hat\rho_{\mathrm{GGE}}\hat n_k\right]$. Evaluating both sides then leads to
\begin{equation}
\lambda_k= -\ln\left(\frac{1}{\braket{\hat n_k}_0}+1\right),
\label{eq:GGE_def}
\end{equation}
with $\braket{\hat n_k}_0=\cosh(2\Theta_k)\sum_{i,j}U_{ik}U_{jk}\braket{\hat a_i^\dagger \hat a_j}_0 +\sinh^2(\Theta_k)$.

\paragraph*{FDRs in integrable real-time dynamics.--} Here we show explicitly that the GGE-FDR in Eq.~\eqref{eq: FDR_dk} emerges  in the non-equilibrium dynamics under the Hamiltonian in Eq.~\eqref{eq:Ham_SW_diag}. In the Heisenberg picture, the rotated operators $\hat d_k$ evolve according to $\hat{d}_k(t)= e^{i\hat H t}\hat{d}_k e^{-i\hat H t}=e^{-i\epsilon_k t}\hat d_k$. Hence, it follows for the two-time correlation functions
\begin{align}
\tilde F&=\frac{1}{2}\braket{\left\lbrace\hat d_k(t_1),\hat d_k^\dagger(t_2) \right\rbrace}_0=e^{-i\epsilon_k (t_1-t_2)} \left(\frac{1}{2}+\braket{\hat{n}_k}_0 \right), \\
\rho&= \braket{\left[ \hat d_k(t_1),\hat d_k^\dagger(t_2)\right]}_0=e^{-i\epsilon_k (t_1-t_2)},
\end{align}
which explicitly shows that Eq.~\eqref{eq: FDR_dk} is fulfilled for all times $t_1,t_2$ as $\braket{\hat{n}_k}_0=\braket{\hat{n}_k}_\mathrm{GGE}=n_{\lambda_k}$ by definition of the GGE. In the sense of the FDR, this integrable model is therefore instantly thermalized to the GGE. 

Similarly, one can also calculate the two-time correlation functions of the number operator $\hat n_k$, which are commuting constants of motion and hence $\rho=\braket{[\hat n_k (t_1),\hat n_{k'}(t_2)]}_0=\braket{[\hat n_k,\hat n_{k'}]}_0=0$. However,  $\tilde F=\braket{\lbrace\hat n_k (t_1),\hat n_{k'}(t_2)\rbrace}_0-\braket{\hat n_k (t_1)}\braket{\hat n_{k'}(t_2)} = \braket{\hat n_k\hat n_{k'}}_0-\braket{\hat n_k}\braket{\hat n_{k'}}\neq 0$ in general and hence there is only an FDR in the sense $ \rho/\tilde F=0$.

\paragraph*{GGE FDR in experimentally observable operators.--} Here we show that a GGE FDR is also obtained for the experimentally accessible operators $\hat a_k$. First of all, we note that
\begin{equation}
\hat a_k(t)=\cosh(\Theta_k)e^{i\epsilon_k t} \hat d_k-\sinh(\Theta_k)e^{-i\epsilon_k t} \hat d_k^\dagger,
\end{equation}
from which it follows that
\begin{align}
\tilde F&=\frac{1}{2}\braket{\left\lbrace \hat a_k(t_1), \hat a^\dagger_k(t_2)\right\rbrace}\\
&=\frac{1}{2}\left(\cosh^2(\Theta_k) e^{-i\epsilon_k (t_1-t_2)}+\sinh^2(\Theta_k) e^{i\epsilon_k (t_1-t_2)}\right)\notag\\&\qquad\times\left(1+2\braket{\hat{d}_k^\dagger\hat{d}_k}_0\right)\notag\\&\quad-\sinh(\Theta_k)\cosh(\Theta_k)\notag\\&\qquad\times\left(e^{-i\epsilon_k (t_1+t_2)}\braket{\hat{d}_k\hat{d}_k}_0+e^{i\epsilon_k (t_1+t_2)}\braket{\hat{d}_k^\dagger\hat{d}^\dagger_k}_0\right),\\
\rho&=\cosh^2(\Theta_k)e^{-i\epsilon_k (t_1-t_2)}-\sinh^2(\Theta_k)e^{i\epsilon_k (t_1-t_2)},
\end{align}
where one can show that $\braket{\hat{d}_k\hat{d}_k}_0=\braket{\hat{d}_k^\dagger\hat{d}_k^\dagger}_0=\cosh(\Theta_k)\sinh(\Theta_k)(1+2\braket{\hat a_k^\dagger \hat a_k}_0)$. 

In the limit where the central time $T=\frac{1}{2}(t_1+t_2)$ is large, we can apply the rotating wave approximation and neglect the fast rotating terms in $\tilde F$. Fourier transforming with respect to the relative time $t_1-t_2$, we find
\begin{align}
&\rho(\omega)=2\pi\big(\cosh^2(\Theta_k)\delta(\omega-\epsilon_k)-\sinh^2(\Theta_k)\delta(\omega+\epsilon_k)\big),\\
&\tilde F(\omega,T\rightarrow\infty)=2\pi\left(\frac{1}{2}+\braket{\hat d_k^\dagger \hat d_k}_0\right)\times\notag\\&\quad\big(\cosh^2(\Theta_k)\delta(\omega-\epsilon_k)+\sinh^2(\Theta_k)\delta(\omega+\epsilon_k)\big).
\end{align}

Therefore, we can read off the GGE parameter $\lambda_k$ from the peak at $\omega=\epsilon_k$ by
\begin{equation}
\lambda_k=\mathrm{ln}\left(\frac{1}{\frac{\tilde F(\omega=\epsilon_k)}{\rho(\omega=\epsilon_k)}-\frac{1}{2}}+1\right).
\end{equation}
This procedure also works if observation or coherence times are finite and so the $\delta$-peaks are broadened as the peaks in both $\tilde F$ and $\rho$ get broadened equally with the area under the curves staying constant.

The dispersion of the diagonalized Hamiltonian $\epsilon_k$ can be read off from the position of the peaks in $\rho$, whereas the ratio of the two peak heights yields the Bogoliubov angle $\Theta_k$. Hence, from a measurement of this two-time function the Hamiltonian can be ``experimentally diagonalized''. Moreover, the two-time functions of the \emph{rotated} degrees of freedom can now be obtained from the unrotated two-time functions via
\begin{align}
&\braket{\hat d_k(t_1)\hat d_k^\dagger (t_2)}\notag\\&=\cosh^2(\Theta_k)\braket{\hat a_k^\dagger(t_1)\hat a_k (t_2)}+\sinh^2(\Theta_k)\braket{\hat a_k(t_1)\hat a_k^\dagger (t_2)}\notag\\
&-\cosh(\Theta_k)\sinh(\Theta_k)\left(\braket{\hat a_k^\dagger(t_1)\hat a_k^\dagger (t_2)}+\braket{\hat a_k(t_1)\hat a_k (t_2)}\right).
\end{align}
This leads to an alternative method to obtain the $\lambda_k$: the FDRs of the \emph{rotated} degrees of freedom can be obtained and the $\lambda_k$ extracted from Eq.~\eqref{eq: FDR_dk}. This alternative procedure has the advantage of only involving the relative time $t_1-t_2$ even when starting from non-equilibrium initial states, such that we can set $t_2=0$, reducing the experimental effort as only one-point functions have to be measured.

\section{Projective measurement protocol for $F$ in Hubbard model simulators\label{app:F_hubb_details}}
In the following, we derive Eq.~\eqref{eq:F_nn} of the main text. After having evolved the initial state $\ket{\Psi(0)}$ under Hamiltonian $\hat H$ for time $t_1$ and subsequently having measured $\hat n_i$ we get for the post-measurement state
\begin{equation}
\ket{\Psi'(t_1)}=\Bigg\{\begin{array}{lr}
\frac{1}{\sqrt{1-\braket{\Psi(t_1)|\hat n_i|\Psi(t_1) }}}
(1-\hat n_i) \ket{\Psi(t_1)}      & \text{for } \ket{0}_{t_1}\\
\frac{1}{\sqrt{\braket{\Psi(t_1)|\hat n_i|\Psi(t_1)}}}
 \hat n_i\ket{\Psi(t_1)}        & \text{for } \ket{1}_{t_1}
        \end{array},
\end{equation}
where $\ket{0}$/$\ket{1}$ denotes having measured occuption zero/one. Subsequently time evolving for time $t_2-t_1$, we find for the final measurement of $\hat n_j$ that
\begin{align}
&\braket{\hat n_j(t_2)}\bigg|_{\ket{0},\ket{1}}\notag=\\&\Bigg\{\begin{array}{lr}
\frac{1}{1-\braket{\hat n_i (t_1)}} \braket{ (1-\hat n_i(t_1)) \hat n_j(t_2)(1-\hat n_i(t_1))}      & \text{for } \ket{0}_{t_1}\\
 \frac{1}{\braket{\hat n_i (t_1)}}  \braket{ \hat n_i(t_1) \hat n_j(t_2)\hat n_i(t_1)}       & \text{for } \ket{1}_{t_1},
        \end{array}
\end{align}
where we switched to the Heisenberg picture. Rearranging terms, one can then deduce Eq.~\eqref{eq:F_nn} of the main text.

\subsubsection*{Non-destructive projective measurement in optical lattices using tweezers}
Here we present several schemes to implement the spatially resolved projective measurement in optical lattices.

\paragraph*{Shining a tweezer on the lattice.--}  Following Ref.~\cite{uhrich_probing_2019}, a tightly focussed tweezer can be used to map the occupation of a site in the 2D optical lattice to the one of the tweezer. Moving the tweezer away from the lattice then makes it possible to measure the occupation without disturbing the rest of the system. For this protocol to work, moving the tweezer should be faster than any time scale in the many body system, especially the tunneling. Tunneling times are on the order of ms in optical lattices~\cite{greiner_quantum_2002} which is longer than the typically $100 \mu \rm s$ it takes to move an optical tweezer over the distance of one lattice site~\cite{barredo_atom-by-atom_2016}.

\paragraph*{Bringing a tweezer next to the lattice.--} Alternatively to shining a tweezer directly on the optical lattice, one may bring it close to a given lattice site~\cite{kantian_lattice-assisted_2015}, which induces tunnelling of strength $J_t$ between the tweezer and the site. Writing the state of an atom being in the tweezer as $\ket{t}$, we can write the effective Hamiltonian as $\hat H=J_t(\ket{t}\bra{1}+\ket{1}\bra{t})$, with $\ket{1}$ denoting the site being occupied. Keeping the tweezer for a time $t$ next to the site induces a ``pulse'' 
\begin{equation}
U=\exp(iHt)=\cos(J_t t)\mathbb{1}+i\sin(J_t t)(\ket{t}\bra{1}+\ket{1}\bra{t}).
\end{equation}
Choosing $t=\pi/J_t$ induces a ``$\pi$-pulse'', mapping the occupation of the site to the initially empty tweezer. Here, $J_t$ needs to be much larger than the energy scales in the Bose-Hubbard model, $J_t\gg J,U$, i.e. the distance of the tweezer from the lattice must be smaller than the lattice spacing (although not much smaller due to the exponential dependence of the tunneling amplitude on the distance~\cite{bloch_many-body_2008}).

\section{Two-time correlation functions in exact diagonalization\label{app_ED}}
In order to calculate the correlation functions $F=\frac{1}{2}\left\langle\left\lbrace \hat A(t_1),\hat B(t_2)\right\rbrace\right\rangle$ and $\rho=\left\langle\left[\hat A(t_1),\hat B(t_2)\right]\right\rangle$ in general, we first time evolve the initial state $\ket{\Psi}$ to $\ket{\Psi(t)}=U(t)\ket{\Psi}\equiv\exp(-i\hat Ht)\ket{\Psi}$  for all times $t$ at which the two-time correlation function should be evaluated. Then, we create a set of four states by acting with $\hat A,\hat{B}$ and their Hermitian conjugates onto $\ket{\Psi(t)}$ and evolve them back for every point in time $t$, such that we arrive at
 $\ket{\Psi_{A}(t)}=\hat{A}(t)\ket{\Psi}$,$\ket{\Psi_{A'}(t)}=\hat{A}^\dagger(t)\ket{\Psi}$,$\ket{\Psi_{B}(t)}=\hat{B}(t)\ket{\Psi}$ and $\ket{\Psi_{B'}(t)}=\hat{B}^\dagger(t)\ket{\Psi}$, where $\hat A(t)=U^\dagger(t)\hat AU(t)$.

From these states we can then calculate $F$ and $\rho$ by evaluating
\begin{align}
F(t_1,t_2)&= \frac{1}{2}\left(\braket{\Psi_{A'}(t_1)|\Psi_{B}(t_2)}+\braket{\Psi_{B'}(t_1)|\Psi_{A}(t_2)}\right),\notag\\\rho(t_1,t_2)&=\left(\braket{\Psi_{A'}(t_1)|\Psi_{B}(t_2)}-\braket{\Psi_{B'}(t_1)|\Psi_{A}(t_2)}\right).
\label{eq:ED_Frho}
\end{align}
for all times $t_1$ and $t_2$.

Simplifications occur if $\hat
B^\dagger=A$ such as for creation/annihilation or $\sigma^+,\sigma^-$ operators, and as then only two states have to be evolved. If additionally $\hat A^\dagger=\hat A$, only a single state needs to be evolved and $F$ and $-(i/2)\rho$ correspond to the real/imaginary parts of the correlation function $\braket{\Psi_{A}(t_1)|\Psi_{A}(t_2)}$.
 
\paragraph*{Efficient numerical evaluation.--} Eq.~\eqref{eq:ED_Frho} can be evaluated efficiently by
writing the states $\ket{\Psi_A(t_1))}$ into a matrix $P_A$, where states for different times are the rows of $P_A$. Then, Eq.~\eqref{eq:ED_Frho} can be evaluated by the matrix product as $\braket{\Psi_{A}(t_1)|\Psi_{B}(t_2)} = [P_{A}^*P_B^T]_{t_1t_2}$.

When using full diagonalization, i.e. obtaining the vector of eigenenergies $\mathbf{E}$ and the matrix $U$ with the eigenvectors as its columns, the forward-backward evolution described above can be efficiently obtained by writing the times $t_1$ into a vector $\mathbf{T}$. By repeating the initial state $\mathrm{dim}(\mathbf{T})$ times in a matrix $P_{\mathrm{ini}}$, the time evolved states follow as
\begin{equation}
P_A=U\exp(i\mathbf{E}\otimes\mathbf{T})\odot U^\dagger A U \exp(-i\mathbf{E}\otimes\mathbf{T})\odot U^\dagger P_{\mathrm{ini}},
\end{equation}
where $\odot$ denotes the Hadamard product (element-wise multiplication) and the exponential is understood element-wise.

\section{Details on protocol using statistical correlations between randomized measurements\label{app:rand}}
The proofs of the relations in Eqs.~\eqref{eq:F_noise},\eqref{eq:C_noise} follow straightforwardly from the ones presented in Ref.~\cite{PhysRevX.9.021061} for the OTOC. 

\paragraph*{Proof of Equ.~\eqref{eq:C_noise}.--} In Ref.~\cite{PhysRevX.9.021061} it was shown from the properties of $u$ that $\overline{\braket{\hat A}_u \braket{\hat B}_u}= \frac{1}{\mathcal{N}_Hc} \sum_{\tau\in S_2}\Tr(\tau \hat A \otimes \hat B)$, where $S_n$ is the permutation group on n letters and $c=\mathcal{N}_H+1$. For $n=2$, $S_2=\lbrace \mathbb{1}, \mathrm{SWAP}\rbrace$, where the SWAP operator acts as $\mathrm{SWAP} (\ket{a}\otimes\ket{b})=\ket{b}\otimes \ket{a}$. By acting with $\tau$ to the left when writing out the trace as a sum over basis states and using that $\Tr(\hat A\otimes \hat B)=\Tr(\hat A)\Tr(\hat B)$, it follows that
\begin{equation}
\overline{\braket{\hat A}_u \braket{\hat B}_u}=\frac{1}{\mathcal{N}_Hc} \left(\Tr(\hat A)\Tr(\hat B)+\Tr(\hat A\hat B)\right).
\end{equation}
Using that $\hat{A}$,$\hat{B}$ are traceless and inserting $\hat{A}\rightarrow \hat{A}(t_1)$, $\hat{B}\rightarrow \hat{B}(t_2)$ we arrive at Eq.~\eqref{eq:C_noise}, where we assumed $\mathcal{N}_H \gg 1$.

\paragraph*{Proof of Equ.~\eqref{eq:F_noise}.--} Similarly, it was shown in Ref.~\cite{PhysRevX.9.021061} that $\overline{\braket{\hat A}_u \braket{\hat B}_u \braket{\hat C}_u}= \frac{1}{c'} \sum_{\tau\in S_3}\Tr(\tau \hat A \otimes \hat B \otimes \hat C)$, where $c'=\mathcal{N}_H(\mathcal{N}_H+1)(\mathcal{N}_H+2)$. Summing over all possible permutations $\tau$, inserting $\hat C=\rho_0$, $\Tr{\rho_0} =1$ and $\hat{A}\rightarrow \hat{A}(t_1)$, $\hat{B}\rightarrow \hat{B}(t_2)$ we get
\begin{align}
&\overline{\braket{\hat A(t_1)}_u \braket{\hat B(t_2)}_u \braket{\rho_0}_u}\notag\\&=\frac{1}{c'}\bigg(\Tr{\hat A}\Tr{\hat B}+\Tr{\hat A}\Tr{(\rho_0 \hat B(t_2))}+\Tr{\hat B}\Tr{(\rho_0 \hat A(t_1))}\notag\\
&+\Tr{(\hat A(t_1) \hat B(t_2))}+\Tr{\rho_0 \hat A(t_1)\hat B(t_2)}+\Tr{\rho_0 \hat B(t_1)\hat A(t_2)}\bigg).
\end{align}
Assuming that the terms in the first row vanish for traceless $\hat{A}$,$\hat{B}$, we arrive at Eq.~\eqref{eq:F_noise}, where we assumed $\mathcal{N}_H \gg 1$.

\paragraph*{Special case: Thermal equilibrium.--} The above protocol can also be used to measure the equilibrium structure factor $F(t_1-t_2)$ by inserting $\rho_0=\rho_\beta=(1/Z)e^{-\beta \hat H}$, which via the FDR then yields the equilibrium spectral function of the operators $\hat A$ and $\hat B$. In cold atom experiments, this protocol may be used to obtain the density-density (particle-hole) spectral function for $\hat{A}= \hat B=\hat n$. For platforms in which it is difficult to prepare thermal states, but moments of the many-body Hamiltonian can be measured (such as trapped ions), finite temperature spectral functions may still be measured in a high temperature expansion~\cite{PhysRevX.9.021061}.

\section{Higher order time ordered correlation functions from Ramsey pulses and projective measurements\label{eq:threepoint}}
\begin{figure}
\includegraphics{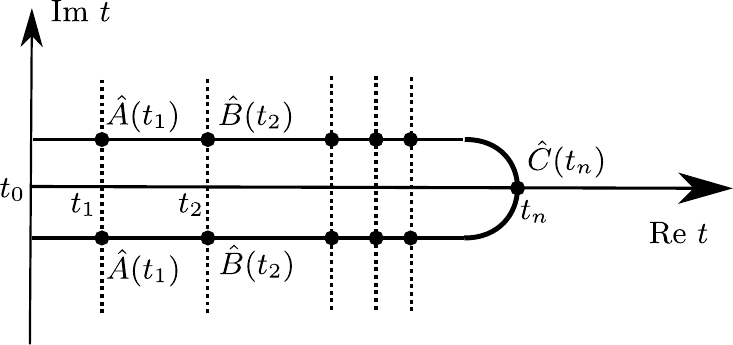}
\caption{\textbf{Closed time contour depiction of the subclass of $(2n+1)$ point correlation functions accessible via protocols with $n$ $\pi$-pulses.--} Starting from the initial time $t_0$, operators are inserted along the contour at times $t_1,t_2,\dots,t_{n}$ by local pulses. At $t_{n+1}$ the operator $\hat{C}$ gets measured and the evolution is stopped. The operator measured by the protocol can then be obtained by starting from $t_0$ on the upper branch up to $t_{n+1}$ and then backward on the lower branch back to $t_0$, i.e. $\braket{\hat A(t_1)\hat{B}(t_2)\cdots \hat{C}(t_{n+1})\cdots \hat{B}(t_2)\hat{A}(t_1}$. Note that this is only a subclass of the correlation functions obtainable by the protocols presented in this section and all operators are given by Pauli matrices. \label{fig:corrfcts_contour}}
\end{figure}

Here we show how to generalize the projective measurement/Ramsey protocols to measure higher order time-ordered correlation functions by using more than one pulse/projection before the final measurement. Here we present the case for two pulses and two projections. We show that  from this sequence $\emph{all}$ three point time ordered correlation functions can be obtained. These are given by the nested (anti-) commutators $\braket{\lbrace A(t_1),\lbrace B(t_2), C(t_3)\rbrace\rbrace}$, $\braket{\lbrace A(t_1),[ B(t_2), C(t_3)]\rbrace}$,$\braket{[ A(t_1),\lbrace B(t_2), C(t_3)\rbrace]}$ and $\braket{[ A(t_1),[ B(t_2), C(t_3)]]}$. The appearance of a anti-/commutator is obtained by a projection/pulse, respectively. 

Apart from all three point correlators, also a subclass of four point and five point functions can be obtained from the two pulse/projection protocol. Furthermore, we show for arbitrary $n$ that a particular ($2n+1$)-point correlation function can be obtained from an $n$ pulse sequence.
 
\paragraph*{Two pulses.--} By using a two-pulse generalization of the commutator protocol discussed in the main text, i.e. evolve until time $t_1$, apply local rotation $\hat R_i^\alpha(\theta)$, evolve until time $(t_2-t_1)$, apply a local rotation $\hat R^\gamma_k(\theta)$, evolve until time $(t_2-t_3)$ and finally measure $\hat \sigma_j^\beta$, one can show that
\begin{align}
&\frac{1}{2}\left(\braket{\hat \sigma_j^\beta}_{\theta}+\braket{\hat \sigma_j^\beta}_ {-\theta}\right)=\cos^4\left(\frac{\theta}{2}\right)\braket{\hat\sigma_j^\beta(t_3)}\notag\\&\qquad-\sin^2\left(\frac{\theta}{2}\right)\cos^2\left(\frac{\theta}{2}\right)\big(\braket{[\hat{\sigma}_i^\alpha(t_1),[\hat{\sigma}_k^\gamma(t_2),\hat{\sigma}_j^\beta(t_3)]]}\notag\\
&\qquad
-\braket{\hat{\sigma}_i^\alpha(t_1)\hat{\sigma}_j^\beta(t_3)\hat{\sigma}_i^\alpha(t_1)}-\braket{\hat{\sigma}_k^\gamma(t_2)\hat{\sigma}_j^\beta(t_3)\hat{\sigma}_k^\gamma(t_2)}\big)\notag\\&\qquad
+\sin^4\left(\frac{\theta}{2}\right)
\braket{\hat{\sigma}_i^\alpha(t_1)\hat{\sigma}_k^\gamma(t_2)\hat{\sigma}_j^\beta(t_3)\hat{\sigma}_k^\gamma(t_2)\hat{\sigma}_i^\alpha(t_1)},
\end{align}
which can be used to extract a five-point function of the form depicted in Fig.~\ref{fig:corrfcts_contour} by using $\theta=\pi$. The knowledge of this five-point-function as well as the one point function and the part of the three point correlation function obtainable from the one pulse commutator protocol can then be used to extract the nested commutator in the second row. Similarly, a nested four-point commutator may be obtained by noting that
\begin{align}
&\frac{1}{2}\left(\braket{\hat \sigma_j^\beta}_{\theta}-\braket{\hat \sigma_j^\beta}_ {-\theta}\right)\notag\\&=i\sin\left(\frac{\theta}{2}\right)\cos^3\left(\frac{\theta}{2}\right)\big(\braket{[\hat{\sigma}_i^\alpha(t_1),\hat{\sigma}_j^\beta(t_3)]}\notag\\&\qquad\qquad\qquad\qquad\qquad\qquad+\braket{[\hat{\sigma}_k^\gamma(t_2),\hat{\sigma}_j^\beta(t_3)]}\big)\notag\\
&+i\sin^3\left(\frac{\theta}{2}\right)\cos\left(\frac{\theta}{2}\right)
\big(\braket{[\hat{\sigma}_i^\alpha(t_1),\hat{\sigma}_k^\gamma(t_2)\hat{\sigma}_j^\beta(t_3)\hat{\sigma}_k^\gamma(t_2)]}\notag\\&+\braket{\hat{\sigma}_i^\alpha(t_1)[\hat{\sigma}_k^\gamma(t_2),\hat{\sigma}_j^\beta(t_3)]\hat{\sigma}_i^\alpha(t_2)}\big),
\end{align}
which is however only a subclass of all possible four-point nested commutators (with others expected to appear with a higher number of pulses).

\paragraph*{n pulses.--} While the exact structure of the obtained commutators for arbitrary rotation angles $\theta$ is difficult to obtain for the general case of $n$ pulses, it can be seen that 
\begin{align}
&\frac{1}{2}\left(\braket{\hat \sigma_j^\beta}_{\theta=\pi}+\braket{\hat \sigma_j^\beta}_ {\theta=-\pi}\right)=\braket{\hat \sigma_j^\beta}_{\theta=\pi}\notag\\&=\braket{\hat{\sigma}_i^\alpha(t_1)\hat{\sigma}_k^\gamma(t_2)\cdots\hat{\sigma}_j^\beta(t_{n+1})\cdots\hat{\sigma}_k^\gamma(t_2)\hat{\sigma}_i^\alpha(t_1)},
\end{align}
where $t_{n+1}$ is the time of the measurement after $n$ pulses at times $t_n$. This $(2n+1)$-point-correlation function can be visualized on the closed time contour, see Fig.~\ref{fig:corrfcts_contour}.

\paragraph*{Two projections.--} The same argumentation can be repeated for the case when pulses are replaced by projections, which in general leads to a replacement of commutators with anticommutators. More specifically, for the case of two projections, we get with analogous notation to the two-pulse case
\begin{align}
&\bra{\Psi(t_1)}\hat{P}_i^{+\alpha}\ket{\Psi(t_1)}\bra{\Psi(t_2)}\hat{P}_k^{+\alpha}\ket{\Psi(t_2)}\braket{\hat{\sigma}_j^\beta(t_2)}_{+\alpha}\notag\\&-\bra{\Psi(t_1)}\hat{P}_i^{-\alpha}\ket{\Psi(t_1)}\bra{\Psi(t_2)}\hat{P}_k^{-\alpha}\ket{\Psi(t_2)}\braket{\hat{\sigma}_j^\beta(t_3)}_{-\alpha}\notag\\
&=\frac{1}{8}
\bigg(\braket{\lbrace\hat{\sigma}_j^\beta(t_3),\hat{\sigma}_k^\gamma(t_2)+\hat{\sigma}_i^\alpha(t_1)\rbrace}\notag\\&+\braket{\hat{\sigma}_i^\alpha(t_1)\lbrace\hat{\sigma}_k^\gamma(t_2),\hat{\sigma}_j^\beta(t_3)\rbrace\hat{\sigma}_i^\alpha(t_1)}\notag\\&+\braket{\lbrace\hat{\sigma}_i^\alpha(t_1),\hat{\sigma}_k^\gamma(t_2)\hat{\sigma}_j^\beta(t_3)\hat{\sigma}_k^\gamma(t_3)\rbrace}\bigg),
\end{align}
and
\begin{align}
&\bra{\Psi(t_1)}\hat{P}_i^{+\alpha}\ket{\Psi(t_1)}\bra{\Psi(t_2)}\hat{P}_k^{+\alpha}\ket{\Psi(t_2)}\braket{\hat{\sigma}_j^\beta(t_2)}_{+\alpha}\notag\\&+\bra{\Psi(t_1)}\hat{P}_i^{-\alpha}\ket{\Psi(t_1)}\bra{\Psi(t_2)}\hat{P}_k^{-\alpha}\ket{\Psi(t_2)}\braket{\hat{\sigma}_j^\beta(t_3)}_{-\alpha}\notag\\
&=\frac{1}{8}\bigg(\braket{\hat\sigma_j(t_3)}+\braket{\lbrace\hat{\sigma}_i^\alpha(t_1),\lbrace\hat{\sigma}_k^\gamma(t_2),\hat{\sigma}_j^\beta(t_3)\rbrace\rbrace}\notag\\&
+\braket{\hat{\sigma}_i^\alpha(t_1)\hat{\sigma}_j^\beta(t_3)\hat{\sigma}_i^\alpha(t_1)}\notag\\&+\braket{\hat{\sigma}_k^\gamma(t_2)\hat{\sigma}_j^\beta(t_3)\hat{\sigma}_k^\gamma(t_2)}\notag\\&\qquad
+\braket{\hat{\sigma}_i^\alpha(t_1)\hat{\sigma}_k^\gamma(t_2)\hat{\sigma}_j^\beta(t_3)\hat{\sigma}_k^\gamma(t_2)\hat{\sigma}_i^\alpha(t_1)}\bigg),
\end{align}
which indeed are the analogous expressions to the two pulse case with commutators replaced by anticommutators. In particular, the nested double anticommutator three point function can be obtained from the last equation.

\paragraph*{Projection followed by pulse.--} A projection at time $t_1$ can also be followed by a pulse at time $t_2$. Different linear combinations of the expectation value of $\hat \sigma_j^\beta(t_3)$ for $\pm \alpha$ and $\pm \theta$ give access to different correlation functions. Here we only note that a nested anticommutator/commutator three point function can be obtained by
\begin{align}
&\bra{\Psi(t_1)}\hat{P}_i^{+\alpha}\ket{\Psi(t_1)}\braket{\hat{\sigma}_j^\beta(t_2)}_{+\alpha,\theta=\pi/4}\notag\\&+\bra{\Psi(t_1)}\hat{P}_i^{-\alpha}\ket{\Psi(t_1)}\braket{\hat{\sigma}_j^\beta(t_3)}_{-\alpha,\theta=-\pi/4}\notag\\&=\frac{1}{4}\braket{\lbrace\hat{\sigma}_i^\alpha(t_1),[\hat{\sigma}_k^\gamma(t_2),\hat{\sigma}_j^\beta(t_3)]\rbrace}.
\end{align} 

\paragraph*{Pulse followed by projection.--} Similarly, if a pulse at time $t_1$ is followed by a projection at time $t_2$,we get
\begin{align}
&\bra{\Psi(t_1)}\hat{P}_i^{+\alpha}\ket{\Psi(t_1)}\braket{\hat{\sigma}_j^\beta(t_2)}_{+\alpha,\theta=\pi/4}\notag\\&+\bra{\Psi(t_1)}\hat{P}_i^{-\alpha}\ket{\Psi(t_1)}\braket{\hat{\sigma}_j^\beta(t_3)}_{-\alpha,\theta=-\pi/4}\notag\\&=\frac{1}{4}\braket{[\hat{\sigma}_i^\alpha(t_1),\lbrace\hat{\sigma}_k^\gamma(t_2),\hat{\sigma}_j^\beta(t_3)\rbrace]},
\end{align}
i.e. commutator and anticommutator are exchanged compared to projection and pulse being in reverse order.

We hence showed that all possible combinations of (anti-)commutator nestings are measurable on the level of three point functions, which means that the complete time ordered three point function can be reconstruncted. Furthermore, we saw that a projector/commutator always leads to an (anti-)commutator. We therefore expect that the structure remains for higher order correlation functions such that all possible (anti-)commutator nestings can be obtained by appropriate combinations of pulses and projections and hence all time ordered $n$ point correlation functions can be accessed.
\bibliography{bib_fdr}
\end{document}